\documentclass[10pt, prb, aps, twocolumn, showpacs, citeautoscript, floatfix, reprint, amsmath, amssymb, notitlepage, superscriptaddress, longbibliography]{revtex4-1}

\usepackage{graphicx}
\usepackage{stmaryrd}
\usepackage{rotating}
\usepackage{amsmath}
\usepackage{amsfonts}
\usepackage{amssymb}
\usepackage{wasysym}
\usepackage[countmax]{subfloat}
\usepackage{dcolumn} % align table columns on decimal point
\usepackage{bm} % bold math
\usepackage{color}

\usepackage{float}
\usepackage{latexsym}

\begin{document}

\title{Dielectric and optical markers originating from quantum geometry}

%\title{Dielectric and optical properties originated from quantum geometry}

%\title{Dielectric and optical properties of insulators caused by quantum metric}

%\title{Local dielectric and optical markers based on quantum geometry}

%\title{Local markers for dielectric and optical properties based on quantum geometry}

%\title{Dielectric and optical properties of insulators determined by quantum metric}

%\title{Local markers for dielectric and optical properties of semiconductors and insulators}

\author{Wei Chen}
\affiliation{Department of Physics, PUC-Rio, 22451-900 Rio de Janeiro, Brazil}

\date{\rm\today}

\begin{abstract}

We elaborate that many non-excitonic dielectric and optical properties of semiconductors and insulators caused by interband absorption are originated from quantum geometry, including charge susceptibility, relative dielectric constant, optical conductivity, dielectric function, refractive index, absorption coefficient, reflectance, and transmittance. The key to this recognition is the complex optical conductivity, which contains the quantum metric in the optical transition matrix element, and the fact that all these dielectric and optical properties can be expressed in terms of the real and imaginary parts of optical conductivity. Our formalism allows to map all these properties to real space lattice sites as local markers, which can help to explain the spatial inhomogeneity of optical properties detected by near-field scanning optical microscopy, as demonstrated by a minimal model of 3D topological insulators. In addition, the dielectric function marker can be used to detect a recently proposed fidelity marker, or equivalently the spread of valence-band Wannier functions generalized to disordered insulators.

\end{abstract}

\maketitle

\section{Introduction}

The recent surge of interest on the notion of quantum metric has triggered a fair amount of search on the physical phenomena related to this quantity. Starting from the fully antisymmetric valence band state of an insulating material, the quantum metric quantifies the overlap of two such states at neighboring momenta\cite{Provost80}. As far as physical measurables are concerned, optical phenomena seem to be the most relevant to to this quantity, since the interband optical transition matrix element is given by the quantum metric\cite{Ozawa18,vonGersdorff21_metric_curvature,Ghosh24,Ezawa24}. Based on this observation, several dielectric and optical phenomena caused by interband transition have been linked to the quantum metric, including the relative dielectric constant that determines the capacitance\cite{Komissarov24}, the real part of optical conductivity that determines the opacity of 2D materials\cite{deSousa23_fidelity_marker}, and the imaginary part of the dielectric function of 3D insulators\cite{CardenasCastillo24_spread_Wannier}.

In this paper, we elaborate that besides the dielectric constant and real part of optical conductivity, many other non-excitonic dielectric and optical properties of semiconductors and insulators are also determined by the quantum metric. This statement is made because all the optical properties like complex charge susceptibility, complex dielectric function, complex refractive index, absorption coefficient, reflectance, and transmittance are all determined by the complex optical conductivity\cite{Fox10,Tanner19}, and the quantum metric enters as the interband optical transition matrix element in both the real and imaginary parts of the optical conductivity at either zero or finite frequency. As a result, the quantum geometric origin of the dielectric and optical properties of insulating materials is evident. This discovery implies that quantum geometry influences the optical properties of these materials even in the macroscopic scale that are perceivable by human eyes, such as the bending of light at material interfaces due to refractive index.

The connection to quantum geometry also allows all these dielectric and optical properties to be defined locally on lattice sites. Utilizing the formalism that maps the topological order and momentum-integrated quantum metric to real space\cite{Bianco11,Prodan10,Prodan10_2,Prodan11,Chen23_universal_marker,Marrazzo19,deSousa23_fidelity_marker}, we show that these properties can be calculated directly from lattice eigenstates and be defined on each lattice site, leading to what we call dielectric and optical markers. These markers can investigate the influence of real space inhomogeneity on the propagation of electromagnetic wave in the nanometer scale, as demonstrated for a minimal model\cite{Zhang09,Liu10} of prototype 3D topological insulators (TIs) that are known to exhibit nontrivial quantum geometry\cite{Matsuura10,Mera22,vonGersdorff21_metric_curvature}. Although these low energy effective models containing only a subset of bands\cite{Schnyder08,Ryu10,Chiu16} may not be sufficient to reproduce the absolute scale of these dielectric and optical properties measured in experiments, they serve our purpose of demonstrating these markers. Particularly for the dielectric function, we will elaborate that the frequency-integration of the dielectric function marker gives so-called spread of valence band Wannier functions generalized to disordered insulators, thereby quantifying the spread under the influence of disorder.

%Besides, these markers can also answer whether the spatial average of these properties is altered by the impurities, similar to the same question for topological markers\cite{Oliveira24_impurity_marker}. 

%On the other hand, through investigating this minimal model, we also realize that to accurately capture the dielectric and optical properties, one must first correctly describe the band structure and the Bloch wave function, e.g., by first-principle calculations or more sophisticated tight-binding models, such that the interband transitions between all the bands can be correctly included.  

Note that impurities are known to be tremendously important to the optical properties of semiconductors. For instance, doping optically active atoms into semiconductors has been a key mechanism for solid-state lasers\cite{Fox10}. Besides, doping can also change the color of the semiconductor. For instance, introducing Cr$^{3+}$ ions into colorless Al$_{2}$O$_{3}$ is the reason that rubies have their charming red color\cite{Fox10}. Conventionally, the effect of disorder or doping is investigated by linear response theory using Green's function in momentum space, or a semiclassical theory that solves for the local density of states under the influence of smooth impurity potential\cite{Shklovskii84}. Another approach is the percolation theory based on a network of resistors that are randomly removed to simulate the binary mixing\cite{Kirkpatrick73}. Alternatively, one may employ phenomenological methods like effective medium approximation and the classical mixing approach based on the Maxwell-Garnett rule, which treat the dielectric properties phenomenologically as simply proportional to the fraction of the two mixing materials\cite{Stroud98,Sihvola99,Tanner19}. However, none of these approaches are truly microscopic since their length scale is far beyond the lattice constant. In comparison, our dielectric and optical markers are truly microscopic since it starts from the Hamiltonian locally perturbed by the disorder and uses the resulting wave function to calculate them, enabling the quantum effects on these doping-induced dielectric and optical properties to be calculated, which should be particularly useful to investigate atomic scale inhomogeneity.

%, and they do not take into account the effect of local wave function in disordered systems, such as quantum interference caused by many-impurities and the local modification of Hamiltonian

%Moreover, the markers help to understand how impurities can help to engineer the dielectric and optical properties, such as enhancing the capacitance and reflectance, that can have a significant technological impact on device engineering.  

%Electromagnetism is local. What this means is that the electric field ${\bf E}$, electric displacement ${\bf D}$, magnetic flux density ${\bf B}$ and magnetic field strength ${\bf H}$ in a dielectric obey Maxwell's equations even down to the nanometer scale. Thus the charge and magnetic susceptibilities can be introduced on individual lattice sites. 

\section{Optical properties and quantum geometry}

\subsection{Optical properties of dielectrics \label{sec:optical_properties}}

Since we aim to elaborate the relation between the non-excitonic optical properties and the quantum geometry, we first give a review on these properties. Although the relations between these properties are well-known\cite{Fox10,Tanner19}, our aim is to elaborate that all of them can be in terms of the complex optical conductivity. Consider the propagation of electromagnetic wave in a 3D dielectric described by the wave equation
\begin{eqnarray}
&&
-{\nabla}^{2}{\bf E}=\frac{\omega^{2}}{c^{2}}\varepsilon(\omega){\bf E}=\frac{\omega^{2}}{c^{2}}\tilde{n}^{2}{\bf E},
\label{wave_eq_in_dielectric}
\end{eqnarray}
which defines the complex dielectric function $\varepsilon(\omega)$ and the complex refractive index $\tilde{n}$. There are two different interpretations for the complex dielectric function. The first interpretation is to consider the change of charge polarization as a current and express $\varepsilon(\omega)$ entirely in terms of the complex optical conductivity $\sigma(\omega)$, and the second is to regard the optical current as a charge polarization times frequency, which amounts to express the dielectric function entirely in terms of the complex charge susceptibility $\chi(\omega)$. Putting these two schemes together, one writes
\begin{eqnarray}
\boldsymbol\varepsilon(\omega)=1
+i\frac{\sigma(\omega)}{\varepsilon_{0}\omega}=1+\chi(\omega).
\label{dielectric_fn_tensor_form}
\end{eqnarray}
The real and imaginary parts of $\tilde{n}(\omega)$ are usually denoted by
\begin{eqnarray}
\tilde{n}=n+i\kappa.
\end{eqnarray}
For an electromagnetic wave propagating along the ${\hat {\bf z}}$-direction ${\bf E}={\bf E}_{0}e^{i(kz-\omega t)}$, Eq.~(\ref{dielectric_fn_tensor_form}) gives $k^{2}=\omega^{2}\tilde{n}^{2}/c^{2}$, and hence $k=\tilde{n}\omega/c=(n+i\kappa)\omega/c$. This leads to a plane wave solution ${\bf E}={\bf E}_{0}e^{-\kappa\omega z/c}e^{i(\omega nz/c-\omega t)}$,
meaning that the speed of light is reduced by $v=c/n$. The imaginary part $\kappa$ causes the wave to attenuate in the media, quantified by considering the intensity of the wave $I\propto {\bf E}^{\ast}\cdot{\bf E}\propto e^{-2\kappa\omega z/c}\equiv e^{-\alpha z}$, leading to the absorption coefficient
\begin{eqnarray}
\alpha=\frac{2\kappa\omega}{c},
\label{absorption_coefficient_definition}
\end{eqnarray}
whose inverse gives the decay length of the attenuation.

Writing the real (subscript 1) and imaginary (subscript 2) parts of the dielectric function, charge susceptibility, and optical conductivity in Eq.~(\ref{dielectric_fn_tensor_form}) as
\begin{eqnarray}
\varepsilon=\varepsilon_{1}+i\varepsilon_{2}=1+\chi_{1}+i\chi_{2},\;\;\;
\sigma=\sigma_{1}+i\sigma_{2},
\end{eqnarray}
one arrives at
\begin{eqnarray}
\varepsilon_{1}=1-\frac{\sigma_{2}}{\varepsilon_{0}\omega},\;\;\;\chi_{1}=-\frac{\sigma_{2}}{\varepsilon_{0}\omega},\;\;\;
\varepsilon_{2}=\chi_{2}=\frac{\sigma_{1}}{\varepsilon_{0}\omega}.
\label{epsilon12_sigma12_relation}
\end{eqnarray}
The real part of dielectric function at zero frequency is also of particular importance, since it represents the relative dielectric constant that determines the capacitance 
\begin{eqnarray}
\varepsilon_{r}=\lim_{\omega\rightarrow 0}\varepsilon_{1}=1+\lim_{\omega\rightarrow 0}\chi_{1}(\omega).
\end{eqnarray}
Furthermore, from $\tilde{n}^{2}=(n+i\kappa)^{2}=\varepsilon=\varepsilon_{1}+i\varepsilon_{2}$, the dielectric function can be expressed in terms of the refractive index as
\begin{eqnarray}
\varepsilon_{1}=n^{2}-\kappa^{2},\;\;\;
\varepsilon_{2}=2n\kappa,
\end{eqnarray}
or conversely,
\begin{eqnarray}
&&\left\{n,\kappa\right\}=\frac{1}{\sqrt{2}}\left[\pm\varepsilon_{1}+(\varepsilon_{1}^{2}+\varepsilon_{2}^{2})^{1/2}\right]^{1/2}
\nonumber \\
&&=\frac{1}{\sqrt{2}}\left\{\pm 1\mp\frac{\sigma_{2}}{\varepsilon_{0}\omega}
+\left[\left(1-\frac{\sigma_{2}}{\varepsilon_{0}\omega}\right)^{2}+\left(\frac{\sigma_{1}}{\varepsilon_{0}\omega}\right)^{2}\right]^{1/2}\right\}^{1/2},
\nonumber \\
\end{eqnarray}
where the upper signs are for $n$ and the lower for $\kappa$. Using these results, the absorption coefficient in Eq.~(\ref{absorption_coefficient_definition}) can also be expressed in terms of optical conductivity
\begin{eqnarray}
\alpha=\frac{\omega\varepsilon_{2}}{cn}=\frac{\sigma_{1}}{\varepsilon_{0}cn}.
\end{eqnarray}
Having $\left\{n,\kappa\right\}$ determined by $\left\{\sigma_{1},\sigma_{2}\right\}$, one can further use the formula for reflectance $R$ and transmittance $T$ exactly at the surface for light incident normally from vacuum to the surface of the material
\begin{eqnarray}
R=\frac{(n-1)^{2}+\kappa^{2}}{(n+1)^{2}+\kappa^{2}},\;\;\;
T=1-R=\frac{4n}{(n+1)^{2}+\kappa^{2}},\;\;\;
\label{reflectance_transmittance_normal_incident}
\end{eqnarray}
to express the reflectance and transmittance in terms of $\left\{\sigma_{1},\sigma_{2}\right\}$. Thus we see that the $\left\{\sigma_{1},\sigma_{2}\right\}$ determines all the dielectric and optical properties. In addition, recall that $\sigma_{2}$ can be obtained from $\sigma_{1}$ via the Kramers-Kronig relation\cite{Tanner19}
\begin{eqnarray}
\sigma_{2}(\omega)=-\frac{2\omega}{\pi}{\cal P}\int_{0}^{\infty}d\omega '\frac{\sigma_{1}(\omega ')}{\omega^{\prime 2}-\omega^{2}}.
\label{Kramers_Kronig_sigma}
\end{eqnarray}
Thus one really only needs the full frequency-dependence of the real part of the optical conductivity $\sigma_{1}(\omega)$ to obtain all the dielectric and optical properties $\left\{\varepsilon_{1},\varepsilon_{2},\chi_{1},\chi_{2},n,\kappa,\alpha,\varepsilon_{r},R,T\right\}$ of an insulating material.

% {\cblue (No, the $T=t^{\ast}t$ is not how they usually define the transmittance. I should say that the transmittance right at the surface is $T=1-R$. Need to correct both my ${\bf k}$-integration result and the optical marker result, and need to correct my book too.)}

%{\cblue (1) Thus this means besides the optical conductivity marker, we also need the relative dielectric constant marker $\varepsilon_{r}({\bf r},\omega)$ which is what Komissarov24 addresses, and is also related to quantum metric. I think the key to their formalism is the linear response theory in real space, since the perturbation is $\delta H=-e{\hat\mu}E^{\mu}$, so one builds a correlator using lattice eigenstates $\langle E_{n}|\hat{\mu}|E_{m}\rangle\langle E_{m}|\hat{\mu}|E_{n}\rangle$, and then work backwards to say that in $k$-space this corresponds to quantum metric $\langle\partial_{\mu}n|m\rangle\langle m|\partial_{\mu}n\rangle$. }

\subsection{Complex optical conductivity caused by quantum metric \label{sec:complex_optical_conductivity}}

In this section, we elaborate how the real and imaginary parts of optical conductivity are both determined by the quantum metric of the valence band state. Consider fully gapped semiconductors and insulators with $N_{-}$ valence bands, then the fully antisymmetric many-body valence band Bloch state at momentum ${\bf k}$ is given by
\begin{eqnarray}
|u^{\rm val}({\bf k})\rangle=\frac{1}{\sqrt{N_{-}!}}\epsilon^{n_{1}n_{2}...n_{N-}}|n_{1}\rangle|n_{2}\rangle...|n_{N_{-}}\rangle.\;\;\;
\label{psi_val}
\end{eqnarray}
The quantum metric $g_{\mu\nu}$ is defined from the overlap of two such states at neighboring momentum\cite{Provost80} 
\begin{eqnarray}
|\langle u^{\rm val}({\bf k})|u^{\rm val}({\bf k+\delta k})\rangle|=1-\frac{1}{2}g_{\mu\nu}({\bf k})\delta k^{\mu}\delta k^{\nu},
\label{uval_gmunu}
\end{eqnarray}
which can be expressed by\cite{Matsuura10,vonGersdorff21_metric_curvature} 
\begin{eqnarray}
&&g_{\mu\nu}({\bf k})=\frac{1}{2}\langle \partial_{\mu}u^{\rm val}|\partial_{\nu}u^{\rm val}\rangle+\frac{1}{2}\langle \partial_{\nu}u^{\rm val}|\partial_{\mu}u^{\rm val}\rangle
\nonumber \\
&&-\langle \partial_{\mu}u^{\rm val}|u^{\rm val}\rangle \langle u^{\rm val}|\partial_{\nu}u^{\rm val}\rangle
\nonumber \\
&&=
\frac{1}{2}\sum_{nm}\left[\langle \partial_{\mu}n|m\rangle\langle m|\partial_{\nu}n\rangle+\langle \partial_{\nu}n|m\rangle\langle m|\partial_{\mu}n\rangle\right],
\nonumber \\
&&\equiv\sum_{nm}g_{\mu\nu}^{nm},
\label{gmunu_T0}
\end{eqnarray}
with $\partial_{\mu}\equiv\partial/\partial k^{\mu}$, and we reserve the index $n$ for valence bands (not to be confused with the refractive index), $m$ for conduction bands, and $\ell$ for all the bands. Furthermore, for the convenience of our discussion, we have labeled each term in the summation by $g_{\mu\nu}^{nm}$ according to the bands involved. Note that $g_{\mu\nu}^{nm}$ is equal to the quantum metric $g_{\mu\nu}$ itself only in two-band systems.

The real part of the optical conductivity $\sigma_{1}({\bf k},\omega)$ at momentum ${\bf k}$ can be readily calculated by linear response theory. We denote ${\hat j}_{\mu}=e\partial_{\mu}H({\bf k})$ with $\partial_{\mu}\equiv\partial/\partial k^{\mu}$ as the current operator along ${\hat{\boldsymbol\mu}}$-direction at momentum ${\bf k}$, $V_{\rm cell}$ as the volume of the unit cell, $f(\varepsilon_{\ell})=1/(e^{\varepsilon_{\ell}/k_{B}T}+1)$ as the Fermi distribution, and $|\ell\rangle$ for eigenstates satisfying $H({\bf k})|\ell\rangle=\varepsilon_{\ell}|\ell\rangle$ with eigenenergies $\varepsilon_{\ell}$. Assuming the longitudinal conductivity is the same in every direction, the real part of the optical conductivity is given by
\begin{eqnarray}
&&\sigma_{1}({\bf k},\omega)=\frac{\pi}{V_{\rm cell}\hbar\omega}\sum_{\ell<\ell '}\langle \ell|{\hat j}_{\mu}|\ell '\rangle\langle \ell '|{\hat j}_{\mu}|\ell\rangle
\nonumber \\
&&\times[f(\varepsilon_{\ell})-f(\varepsilon_{\ell '})]\delta\left(\omega+\frac{\varepsilon_{\ell}}{\hbar}
-\frac{\varepsilon_{\ell '}}{\hbar}\right)
\nonumber \\
&&=\frac{\pi e^{2}\hbar\omega}{V_{\rm cell}}\sum_{\ell<\ell '}\langle \partial_{\mu}\ell|\ell '\rangle\langle \ell '|\partial_{\mu}\ell\rangle
\nonumber \\
&&\times[f(\varepsilon_{\ell})-f(\varepsilon_{\ell '})]\delta\left(\omega+\frac{\varepsilon_{\ell}}{\hbar}
-\frac{\varepsilon_{\ell '}}{\hbar}\right).
\end{eqnarray}
On the other hand, the imaginary part of optical conductivity can be obtained from the real part via the Kramers-Kronig relation in Eq.~(\ref{Kramers_Kronig_sigma}). The optical conductivity measured in real space is then given by the momentum integration
\begin{eqnarray}
\sigma_{i}(\omega)=\int\frac{d^{3}{\bf k}}{\hbar^{3}V_{BZ}}\sigma_{i}({\bf k},\omega),
\end{eqnarray}
which are those entering the expressions of optical properties $\left\{\varepsilon_{1},\varepsilon_{2},\chi_{1},\chi_{2},n,\kappa,\alpha,\varepsilon_{r},R,T\right\}$ in Sec.~\ref{sec:optical_properties}. The above formalism lead to the following expression for the optical conductivity\cite{Yu10}
\begin{eqnarray}
&&\sigma_{1}(\omega)=\frac{\pi e^{2}\omega}{\hbar^{2}}\int\frac{d^{3}{\bf k}}{(2\pi)^{3}}\sum_{\ell<\ell '}\langle \partial_{\mu}\ell|\ell '\rangle\langle \ell '|\partial_{\mu}\ell\rangle
\nonumber \\
&&\times\left[f(\varepsilon_{\ell})-f(\varepsilon_{\ell '})\right]\delta\left(\omega+\frac{\varepsilon_{\ell}}{\hbar}
-\frac{\varepsilon_{\ell '}}{\hbar}\right)
\nonumber \\
&&\equiv\frac{\pi e^{2}}{\hbar^{2}}\,\omega\,{\cal G}_{\mu\mu}^{d}(\omega),
\nonumber \\
&&\sigma_{2}(\omega)=\frac{2e^{2}\omega}{\hbar^{2}}\int\frac{d^{3}{\bf k}}{(2\pi)^{3}}\sum_{\ell<\ell '}\langle \partial_{\mu}\ell|\ell '\rangle\langle \ell '|\partial_{\mu}\ell\rangle
\nonumber \\
&&\times\left[f(\varepsilon_{\ell})-f(\varepsilon_{\ell '})\right]\frac{\varepsilon_{\ell '}/\hbar
-\varepsilon_{\ell}/\hbar}{\omega^{2}-\left(\varepsilon_{\ell '}/\hbar
-\varepsilon_{\ell}/\hbar\right)^{2}}.
\label{sigma1w_sigma2w}
\end{eqnarray}
where ${\cal G}_{\mu\mu}^{d}(\omega)$ is the dressed fidelity number spectral function that has been introduced previously that frequency-integrates to the momentum-integration of quantum metric\cite{deSousa23_fidelity_marker}, whose significance will be discussed in Sec.~\ref{sec:local_theory_insulating_state}. Particularly at zero temperature $T=0$, which is a good approximation at room temperature for the materials with a large band gap $\apprge$eV, one may replace $\ell\rightarrow n$, $\ell '\rightarrow m$, and $f(\varepsilon_{\ell})-f(\varepsilon_{\ell '})\rightarrow 1$ to write 
\begin{eqnarray}
&&\sigma_{1}^{(0)}(\omega)=\frac{\pi e^{2}\omega}{\hbar^{2}}\int\frac{d^{3}{\bf k}}{(2\pi)^{3}}\sum_{nm}g_{\mu\mu}^{nm}
\delta\left(\omega+\frac{\varepsilon_{n}}{\hbar}
-\frac{\varepsilon_{m}}{\hbar}\right),
\nonumber \\
&&\sigma_{2}^{(0)}(\omega)=\frac{2e^{2}\omega}{\hbar^{2}}\int\frac{d^{3}{\bf k}}{(2\pi)^{3}}\sum_{nm}g_{\mu\mu}^{nm}
\frac{\varepsilon_{m}/\hbar
-\varepsilon_{n}/\hbar}{\omega^{2}-\left(\varepsilon_{m}/\hbar
-\varepsilon_{n}/\hbar\right)^{2}},
\nonumber \\
\label{sigma1w_sigma2w_T0}
\end{eqnarray}
from which one sees that both the real and imaginary parts contain the elements of quantum metric $g_{\mu\mu}^{nm}$ that are precisely the optical transition matrix element. Consequently, the $\left\{\varepsilon_{1},\varepsilon_{2},\chi_{1},\chi_{2},n,\kappa,\alpha,\varepsilon_{r},R,T\right\}$ 
calculated from $\left\{\sigma_{1},\sigma_{2}\right\}$ will all contain $g_{\mu\mu}^{nm}$, indicating the quantum geometric origin of all these non-excitonic dielectric and optical properties of semiconductors and insulators. This remarkable feature implies that many optical phenomena that can be perceived in the macroscopic scale by human eyes are caused by quantum geometry, such as the refraction and reflection.

%{\cblue (1) Actually an interesting question is whether the relative dielectric constant $\varepsilon_{r}$ would diverge as the system approaches TPT. I would guess yes, because the quantum metric diverges at the gap-closing point, and the denominator of imaginary part of optical conductivity is just the gap so it also diverges. The question is after momentum integration, would it still diverge, or at least show a cusp wrt tuning parameter. Can make a figure of $\varepsilon_{r}$ versus $M$. }

%Thus we see that the optical transition matrix element in $\sigma_{1}({\bf k},\omega)$ is precisely given by the elements of quantum metric $g_{\mu\mu}^{\ell\ell '}$, indicating the quantum geometric origin of the optical conductivity. In particular, at zero temperature $T\rightarrow 0$, the Fermi factor becomes unity $[f(\varepsilon_{\ell})-f(\varepsilon_{\ell '})]\rightarrow 1$, and one can simply replace $\ell\rightarrow n$ by the valence bands and $\ell '\rightarrow m$ by the conduction bands such that 
%\begin{eqnarray}
%&&\lim_{T\rightarrow 0}\sigma_{1}({\bf k},\omega)=\frac{\pi e^{2}\hbar\omega}{V_{\rm cell}}\sum_{nm}g_{\mu\mu}^{nm}\delta\left(\omega+\frac{\varepsilon_{n}}{\hbar}
%-\frac{\varepsilon_{m}}{\hbar}\right)
%\nonumber \\
%&&=\frac{\pi e^{2}}{V_{\rm cell}}\hbar\omega\,g_{\mu\mu}({\bf k},\omega),
%\end{eqnarray}
%where $g_{\mu\mu}({\bf k},\omega)$ is what we call the quantum metric spectral function that frequency integrates to quantum metric $\int_{0}^{\infty}d\omega\,g_{\mu\mu}({\bf k},\omega)=g_{\mu\mu}({\bf k})$. 

\subsection{Optical markers and correlators \label{sec:optical_markers}}

Owing to the presence of quantum metric as the optical transition matrix element, one can adopt the formalism of topological marker\cite{Bianco11,Prodan10,Prodan10_2,Prodan11,Chen23_universal_marker} and fidelity marker\cite{deSousa23_fidelity_marker} (also called localization marker\cite{Marrazzo19}) to map all these frequency-dependent dielectric and optical properties into real space and directly calculate them from a lattice Hamiltonian, which we formulate below. We observe that the momentum integration of $\sigma_{1}(\omega)$ in Eq.~(\ref{sigma1w_sigma2w}) can be written into real space by\cite{Bianco11,deSousa23_fidelity_marker}
\begin{eqnarray}
&&\int\frac{d^{3}{\bf k}}{(2\pi)^{3}}\sum_{\ell<\ell '}\langle \partial_{\mu}\ell|\ell '\rangle\langle \ell '|\partial_{\mu}\ell\rangle
\nonumber \\
&&\times[f(\varepsilon_{\ell})-f(\varepsilon_{\ell '})]\delta\left(\omega+\frac{\varepsilon_{\ell}}{\hbar}
-\frac{\varepsilon_{\ell '}}{\hbar}\right)
\nonumber \\
&&=\frac{1}{\hbar^{2}}\int\frac{d^{3}{\bf k}}{(2\pi)^{3}}\sum_{\ell<\ell '}\langle\psi_{\ell}^{\bf k}|{\hat\mu}|\psi_{\ell '}^{\bf k}\rangle\langle\psi_{\ell '}^{\bf k}|{\hat\mu}|\psi_{\ell}^{\bf k}\rangle
\nonumber \\
&&\times[f(\varepsilon_{\ell}^{\bf k})-f(\varepsilon_{\ell '}^{\bf k})]\delta\left(\omega+\frac{\varepsilon_{\ell}^{\bf k}}{\hbar}
-\frac{\varepsilon_{\ell '}^{\bf k}}{\hbar}\right)
\nonumber \\
&&=\frac{\hbar}{V_{\rm cell}}\int\frac{d^{3}{\bf k}}{(2\pi\hbar)^{3}/V_{\rm cell}}\int\frac{d^{3}{\bf k '}}{(2\pi\hbar)^{3}/V_{\rm cell}}
\nonumber \\
&&\times\sum_{\ell<\ell '}\langle\psi_{\ell}^{\bf k}|{\hat\mu}|\psi_{\ell '}^{\bf k'}\rangle\langle\psi_{\ell '}^{\bf k'}|{\hat\mu}|\psi_{\ell}^{\bf k}\rangle
\nonumber \\
&&\times[f(\varepsilon_{\ell}^{\bf k})-f(\varepsilon_{\ell '}^{\bf k'})]\delta\left(\omega+\frac{\varepsilon_{\ell}^{\bf k}}{\hbar}
-\frac{\varepsilon_{\ell '}^{\bf k'}}{\hbar}\right)
\nonumber \\
&&=\frac{\hbar}{NV_{\rm cell}}\sum_{\ell<\ell '}\langle E_{\ell}|{\hat \mu}|E_{\ell '}\rangle\langle E_{\ell '}|{\hat \mu}|E_{\ell}\rangle
\nonumber \\
&&\times[f(E_{\ell})-f(E_{\ell '})]\delta\left(\omega+\frac{E_{\ell}}{\hbar}
-\frac{E_{\ell '}}{\hbar}\right),
\label{from_kspace_to_realspace_marker}
\end{eqnarray}
where we have explicitly labeled the momenta ${\bf k}$ and ${\bf k'}$ of the eigenstates $|\psi_{\ell}\rangle$ and eigenenergies $\varepsilon_{\ell}$ in momentum space, and $|E_{\ell}\rangle$ is the eigenstate of a real space lattice Hamiltonian satisfying $H|E_{\ell}\rangle=E_{\ell}|E_{\ell}\rangle$ with eigenenergy $E_{\ell}$. In deriving this expression, we have used\cite{Marzari97,Marzari12} $i\langle\ell '|\partial_{\mu}\ell\rangle=\langle\psi_{\ell '}^{\bf k}|{\hat\mu}|\psi_{\ell}^{\bf k}\rangle/\hbar$, where $\langle{\bf r}|\ell\rangle=e^{-i{\bf k\cdot r}}\langle{\bf r}|\psi_{\ell}^{\bf k}\rangle$ defines the relation between the full Bloch state $|\psi_{\ell}^{\bf k}\rangle$ and its cell-periodic part $|\ell\rangle$, and ${\hat\mu}$ is the position operator (all the degrees of freedom $\gamma$ within the unit cell at ${\bf r}$ are assigned by the same ${\bf r}$). The $\delta$-function in this expression can be simulated in practice by a Lorentzian $\delta(x)=(\eta/\pi)/(x^{2}+\eta^{2})$ with some small broadening $\eta$. Likewisely, the momentum integration in $\sigma_{2}(\omega)$ in Eq.~(\ref{sigma1w_sigma2w}) can be written as 
\begin{eqnarray}
&&\int\frac{d^{3}{\bf k}}{(2\pi)^{3}}\sum_{\ell<\ell '}\langle \partial_{\mu}\ell|\ell '\rangle\langle \ell '|\partial_{\mu}\ell\rangle
\nonumber \\
&&\times\left[f(\varepsilon_{\ell})-f(\varepsilon_{\ell '})\right]\frac{\varepsilon_{\ell '}/\hbar
-\varepsilon_{\ell}/\hbar}{\omega^{2}-\left(\varepsilon_{\ell '}/\hbar
-\varepsilon_{\ell}/\hbar\right)^{2}}
\nonumber \\
&&=\frac{\hbar}{NV_{\rm cell}}\sum_{\ell<\ell '}\langle E_{\ell}|{\hat \mu}|E_{\ell '}\rangle\langle E_{\ell '}|{\hat \mu}|E_{\ell}\rangle
\nonumber \\
&&\times\left[f(E_{\ell})-f(E_{\ell '})\right]\frac{E_{\ell '}/\hbar
-E_{\ell}/\hbar}{\omega^{2}-\left(E_{\ell '}/\hbar
-E_{\ell}/\hbar\right)^{2}},
\end{eqnarray}
Denoting the projector to a specific eigenstate by $S_{\ell}=|E_{\ell}\rangle\langle E_{\ell}|$, the above analysis implies that the average optical conductivity of the whole system can be written as summation over that on each lattice site located at position ${\bf r}$
\begin{eqnarray}
\sigma_{i}(\omega)=\frac{1}{N}\sum_{\bf r}\sigma_{i}({\bf r},\omega),
\end{eqnarray}
which introduces the optical conductivity markers 
\begin{widetext}
\begin{eqnarray}
&&\sigma_{1}({\bf r},\omega)=\left(\frac{\pi e^{2}}{\hbar V_{\rm cell}}\right)\omega\sum_{\gamma}\langle{\bf r},\gamma|\left\{\sum_{\ell<\ell '}S_{\ell}{\hat\mu}S_{\ell '}{\hat\mu}S_{\ell}[f(E_{\ell})-f(E_{\ell '})]\delta\left(\omega+\frac{E_{\ell}}{\hbar}
-\frac{E_{\ell '}}{\hbar}\right)\right\}|{\bf r},\gamma\rangle,
\nonumber \\
&&\sigma_{2}({\bf r},\omega)=\left(\frac{2e^{2}}{\hbar V_{\rm cell}}\right)\omega\sum_{\gamma}\langle{\bf r},\gamma|\left\{\sum_{\ell<\ell '}S_{\ell}{\hat\mu}S_{\ell '}{\hat\mu}S_{\ell}\left[f(E_{\ell})-f(E_{\ell '})\right]\frac{E_{\ell '}/\hbar
-E_{\ell}/\hbar}{\omega^{2}-\left(E_{\ell '}/\hbar
-E_{\ell}/\hbar\right)^{2}}\right\}|{\bf r},\gamma\rangle,
\label{optical_conductivity_markers}
\end{eqnarray}
where $\gamma$ denotes all the internal degrees of freedom inside the unit cell at position ${\bf r}$, such as spins and orbitals. Once these markers are calculated from a lattice Hamiltonian, all the other optical markers ${\cal O}({\bf r},\omega)$ with ${\cal O}=\left\{\varepsilon_{1},\varepsilon_{2},\chi_{1},\chi_{2},n,\kappa,\alpha,\varepsilon_{r},R,T\right\}$ can be obtained from the conversion formulas given in Sec.~\ref{sec:optical_properties}, which remain true even when these quantities are defined locally at ${\bf r}$. We further decompose the optical conductivity markers at ${\bf r}$ to an optical conductivity correlator that accounts for the nonlocal optical responses caused by applying a field at ${\bf r'}$ 
\begin{eqnarray}
&&\sigma_{1}({\bf r,r'},\omega)=\left(\frac{\pi e^{2}}{\hbar V_{\rm cell}}\right)\omega\sum_{\gamma}\langle{\bf r},\gamma|\left\{\sum_{\ell<\ell '}S_{\ell}{\hat\mu}S_{\ell '}{\hat\mu}_{\bf r'}S_{\ell}[f(E_{\ell})-f(E_{\ell '})]\delta\left(\omega+\frac{E_{\ell}}{\hbar}
-\frac{E_{\ell '}}{\hbar}\right)\right\}|{\bf r},\gamma\rangle,
\nonumber \\
&&\sigma_{2}({\bf r,r'},\omega)=\left(\frac{2e^{2}}{\hbar V_{\rm cell}}\right)\omega\sum_{\gamma}\langle{\bf r},\gamma|\left\{\sum_{\ell<\ell '}S_{\ell}{\hat\mu}S_{\ell '}{\hat\mu}_{\bf r'}S_{\ell}\left[f(E_{\ell})-f(E_{\ell '})\right]\frac{E_{\ell '}/\hbar
-E_{\ell}/\hbar}{\omega^{2}-\left(E_{\ell '}/\hbar
-E_{\ell}/\hbar\right)^{2}}\right\}|{\bf r},\gamma\rangle,
\label{nonlocal_optical_conductivity}
\end{eqnarray}
\end{widetext}
where ${\hat\mu}_{\bf r'}$ is the position operator at site ${\bf r'}$. The sum of the optical correlator recovers the optical marker
\begin{eqnarray}
\sigma_{i}({\bf r},\omega)=\sum_{\bf r'}\sigma_{i}({\bf r,r'},\omega),
\end{eqnarray}
since ${\hat\mu}=\sum_{\bf r'}{\hat\mu}_{\bf r'}$. This decomposition is made because one observes that in Eq.~(\ref{from_kspace_to_realspace_marker}), the second position operator comes from the applied field, and therefore one can simply decompose it to account for a local field applied at ${\bf r'}$\cite{Molignini23_Chern_marker}. These correlators can help to understand how the nonlocal responses depend on the distance ${\bf r-r'}$ between the local field and the local current, which deserve to be investigated elsewhere in detail. In the present work, we focus on the local markers ${\cal O}({\bf r},\omega)$. Finally, the numerical calculations of the optical conductivity markers in Eq.~(\ref{optical_conductivity_markers}) and nonlocal optical conductivity in Eq.~(\ref{nonlocal_optical_conductivity}) can be conveniently implimented using a method that simplifies the product of projectors\cite{deSousa23_fidelity_marker}, as elaborated in Appendix \ref{apx:numerical_optical_conductivity_marker}.

Concerning the experimental detection of these markers, we propose the following two scenarios. Firstly, the dielectric function and refractive index in the macroscopic scale are usually measured experimentally by ellipsometry, which detects the rotation of polarization of the light after hitting the surface of a material\cite{Yu10}. The dielectric and optical properties measured in this way correspond to the spatially averaged markers ${\cal O}(\omega)=\sum_{\bf r}{\cal O}({\bf r},\omega)/N$. Thus if the sample contains impurities, one can first investigate the local marker and then spatially average them to compare with the experimental data, allowing the effect of disorder to be investigated.  

Secondly, to detect the atomic scale variation of the marker around the impurities, as will be elaborated in Sec.~\ref{sec:applications}, the most promising technique is the near-field scanning optical microscopy (NSOM)\cite{Betzig86,Hsu01,Kim07}. This experimental technique measures the spatial variation of refractive index and reflectivity down to nanometer scale, offering the possibility of mapping the spatial variation of these markers. However, to our knowledge, the highest resolution of NSOM to date is about few tenth of nanometers\cite{Tranca18}, which has not yet reach atomic scale, and it remains unclear to us whether NSOM can ever reach the atomic scale to measure the markers on each lattice site. Nevertheless, this few tenth of nanometer resolution of NSOM corresponds to averaging the optical markers over few hundred lattice sites, so in principle our markers also allow to describe the inhomogeneity in the NSOM data. Finally, we also remark that transmittance and reflectance are usually derived by matching the electromagnetic wave at the interface\cite{Fox10,Tanner19}, assuming both sides of the interface are continuous media with some uniform values of refractive index $\left\{n,\kappa\right\}$. However, this is no longer valid in the atomic scale, since the transmittance and reflectance can be defined for each atomic layer away from the interface, and one should match the electromagnetic waves layer by layer. Our transmittance and reflectance markers provide exactly the tool to perform this matching.

%The optical correlators for all the other quantities ${\cal A}({\bf r,r'},\omega)$ can then be obtained by inserting $\sigma_{i}({\bf r,r'},\omega)$ everywhere into the conversion formulas given in Sec.~\ref{sec:optical_properties}.

%{\cblue (1) Actually another significance is that this marker not only respects the Maxwell's equations, but also the Bloch theorem of lattice wave function.  }

%{\cblue (1) No, the right logic to get the complex optical conductivity is that the real part $\sigma_{1}$ is obtained by linear response, and then the imaginary $\sigma_{2}$ can be obtained from Kramers-Kronig. See Yu and Cardona book page 261, and Tanner book Sec 10.2. }

\subsection{Implication on the local theory of insulating state \label{sec:local_theory_insulating_state}}

A profund implication of the optical markers is the experimental detection of the so-called local theory of insulating state\cite{Marrazzo19,deSousa23_fidelity_marker}, which generalizes the spread of valence-band Wannier functions in homogeneous insulators to disordered insulators, as we now explain. In homogeneous insulators, the spread is introduced through the Wannier state $|{\bf R}\ell\rangle$ centering around the home cell ${\bf R}$ defined from the cell periodic part of the Bloch state $|\ell\rangle$ of the $\ell$-th band via
\begin{eqnarray}
&&|\ell\rangle=\sum_{{\bf R}}e^{-i {\bf k}\cdot({\hat{\bf r}}-{\bf R})/\hbar}|{\bf R}\ell\rangle,
\nonumber \\
&&|{\bf R} \ell\rangle=\sum_{\bf k}e^{i {\bf k}\cdot({\hat{\bf r}}-{\bf R})/\hbar}|\ell\rangle.
\label{Wannier_basis}
\end{eqnarray}
The spread is defined from the variance of the charge distribution of the valence-band Wannier states\cite{Marzari97,Marzari12,CardenasCastillo24_spread_Wannier}
\begin{eqnarray}
\Omega=\sum_{n}\left[\langle{\bf 0}n|r^{2}|{\bf 0}n\rangle-\langle{\bf 0}n|{\bf r}|{\bf 0}n\rangle^{2}\right]=\Omega_{I}+\tilde{\Omega},
\label{Omega_original}
\end{eqnarray}
which is conventionally separated to a gauge-invariant part $\Omega_{I}$ and a gauge-dependent part $\tilde{\Omega}$. Our focus is the gauge-invariant part known to be given by the momentum-integration of the quantum metric\cite{Marzari97,Souza00}
\begin{eqnarray}
\Omega_{I}=\frac{V_{\rm cell}}{\hbar}\sum_{\mu}\int\frac{d^{3}{\bf k}}{(2\pi)^{3}}\,g_{\mu\mu}({\bf k}).
\label{OmegaI_kintegration_metric}
\end{eqnarray}
Owing to the relation between $\sigma_{1}(\omega)$ and $g_{\mu\mu}^{nm}$ at zero temperature given in Eqs.~(\ref{sigma1w_sigma2w_T0}), and the relation between $\sigma_{1}(\omega)$ and $\varepsilon_{2}(\omega)$ given in Eq.~(\ref{epsilon12_sigma12_relation}), it is recently proposed that the spread can be experimentally extracted from the frequency-integration of the imaginary part of dielectric function $\varepsilon_{2}(\omega)$ by\cite{CardenasCastillo24_spread_Wannier}
\begin{eqnarray}
\Omega_{I}=\frac{V_{\rm cell}\varepsilon_{0}}{\pi e^{2}}\times 3\int_{0}^{\infty}d(\hbar\omega)\,\varepsilon_{2}(\omega),
\label{OmegaI_wintegral_epsilon2}
\end{eqnarray}
allowing to detect $\Omega_{I}$ without any fitting parameters. For instance, using the experimental data\cite{Reed99,Madelung04,Greenaway65} for $\varepsilon_{2}(\omega)$, one extracts $\Omega_{I}=\left\{68.1, 83.24, 407.48\right\}\AA^{2}$ for Si, Ge, and Bi$_{2}$Te$_{3}$, respectively.

%Roughly speaking, the spread represents the range of atomic orbitals that participate in the kinetics (e.g., hopping) of electrons on the lattice. 

However, the above formalism does not apply to arbitrarily disordered insulators since the Bloch state $|\ell\rangle$ is ill-defined, let alone the Wannier state $|{\bf R}\ell\rangle$. Nevertheless, one can still introduce a similar concept through converting the momentum integration of quantum metric in Eq.~(\ref{OmegaI_kintegration_metric}) to a local quantity ${\cal G}_{\mu\mu}({\bf r})$, which has been called the localization marker\cite{Marrazzo19} or fidelity marker\cite{deSousa23_fidelity_marker}, and then define a local spread $\Omega_{I}({\bf r})$ accordingly, yielding
\begin{eqnarray}
&&\Omega_{I}({\bf r})=\frac{V_{\rm cell}}{\hbar}\sum_{\mu}{\cal G}_{\mu\mu}({\bf r})
\nonumber \\
&&=\sum_{\mu}\sum_{\gamma}\langle{\bf r},\gamma|P{\hat\mu}Q{\hat\mu}P|{\bf r},\gamma\rangle,
\end{eqnarray} 
where $P=\sum_{n}|E_{n}\rangle\langle E_{n}|$ and $Q=\sum_{m}|E_{m}\rangle\langle E_{m}|$ are projectors to the filled and empty lattice eigenstates, respectively, that incorporate the effect of disorder. In the homogeneous and thermodynamic limit, $\Omega_{I}({\bf r})$ recovers the $\Omega_{I}$ in Eq.~(\ref{Omega_original}), and hence $\Omega_{I}({\bf r})$ may be regarded as the generalization of $\Omega_{I}$ to disordered insulators. Remarkably, our optical marker formalism in Sec.~\ref{sec:optical_markers} implies that this local spread is equivalently the frequency-integration of the dielectric function marker at zero temperature 
\begin{eqnarray}
\Omega_{I}({\bf r})=\frac{V_{\rm cell}\varepsilon_{0}}{\pi e^{2}}\times 3\int_{0}^{\infty}d(\hbar\omega)\,\varepsilon_{2}({\bf r},\omega)|_{T=0},
\label{OmegaI_wintegral_epsilon2_atr}
\end{eqnarray}
in a way completely analogous to Eq.~(\ref{OmegaI_wintegral_epsilon2}). Thus the local spread $\Omega_{I}({\bf r})$ may be extracted from the experimental measurement of $\varepsilon_{2}({\bf r},\omega)$. In practice, the local spread averaged over several lattice sites $\sum_{\bf r}\Omega_{I}({\bf r})/N$, where $N$ may be the spatial resolution of NSOM or the whole sample, can be detected by the NSOM or ellipsometry, thus quantifying the effect of disorder. For instance, $\Omega_{I}({\bf r})$ can answer whether the spread is robust against disorder, similar to the same question to the topological order\cite{Oliveira24_impurity_marker}. Furthermore, comparing the $\varepsilon_{2}(\omega)$ in disordered and clean semiconductors measured by ellipsometry tells us how the spread, on average, is influenced by the disorder, and can be quantified theoretically by $\Omega_{I}({\bf r})$.

\section{Applications to 3D topological materials \label{sec:applications}}

%{\cblue (1) I cannot get a sensible result for the dielectric function of linear Dirac model, so I decide to give up and use directly the lattice Dirac model to calculate. }

\subsection{Minimal model for 3D TIs \label{sec:3DTI_kspace}}

As a concrete example, we now discuss a minimal model of 3D TIs in class AII that is relevant to real materials such as Bi$_{2}$Se$_{3}$ and Bi$_{2}$Te$_{3}$. The model is described by a $4\times 4$ Dirac Hamiltonian in momentum space $H({\bf k})={\bf d(k)}\cdot{\boldsymbol\Gamma}$, with $\Gamma_{i}$ the Dirac matrices\cite{Zhang09,Liu10}
\begin{eqnarray}
\Gamma_{1\sim 5}=\left\{\sigma^{x}\otimes\tau^{x},\sigma^{y}\otimes\tau^{x},\sigma^{z}\otimes\tau^{x},
I_{\sigma}\otimes\tau^{y},I_{\sigma}\otimes\tau^{z}\right\},
\nonumber \\
\end{eqnarray} 
where the basis is $\left(c_{{\bf k}P1_{-}^{+}\uparrow},c_{{\bf k}P2_{+}^{-}\uparrow},c_{{\bf k}P1_{-}^{+}\downarrow},c_{{\bf k}P2_{+}^{-}\downarrow}\right)^{T}\equiv\left(c_{{\bf k}s\uparrow},c_{{\bf k}p\uparrow},c_{{\bf k}s\downarrow},c_{{\bf k}p\downarrow}\right)^{T}$, with $s$ and $p$ the abbreviations for the $P1_{-}^{+}$ and $P2_{+}^{-}$ orbitals in real materials. The components of the ${\bf d(k)}$-vector are assigned by\cite{Chen20_absence_edge_current}
\begin{eqnarray}
&&d_{5}=M+6B-2B\sum_{i=1}^{3}\cos \left(k_{i}a/\hbar\right),
\nonumber \\
&&d_{1}=A\sin \left(k_{y}a/\hbar\right),\;\;\;
d_{2}=-A\sin \left(k_{x}a/\hbar\right),
\nonumber \\
&&d_{3}=0,\;\;\;
d_{4}=A\sin \left(k_{z}a/\hbar\right).
\label{3D_class_AII_kspace_model}
\end{eqnarray}
The model has $N/2$-fold degenerate valence bands with energy $\varepsilon_{n}=-d$ and conduction bands with energy $\varepsilon_{m}=d$, where $d=\sqrt{\sum_{i=0}^{3}d_{i}^{2}}$. This minimal model captures the topological phases of these materials caused by the inversion of low energy bands, but it also omits other bands that are crucial to the dielectric and optical properties, as we shall see below.

\begin{figure}[ht]
\begin{center}
\includegraphics[clip=true,width=0.99\columnwidth]{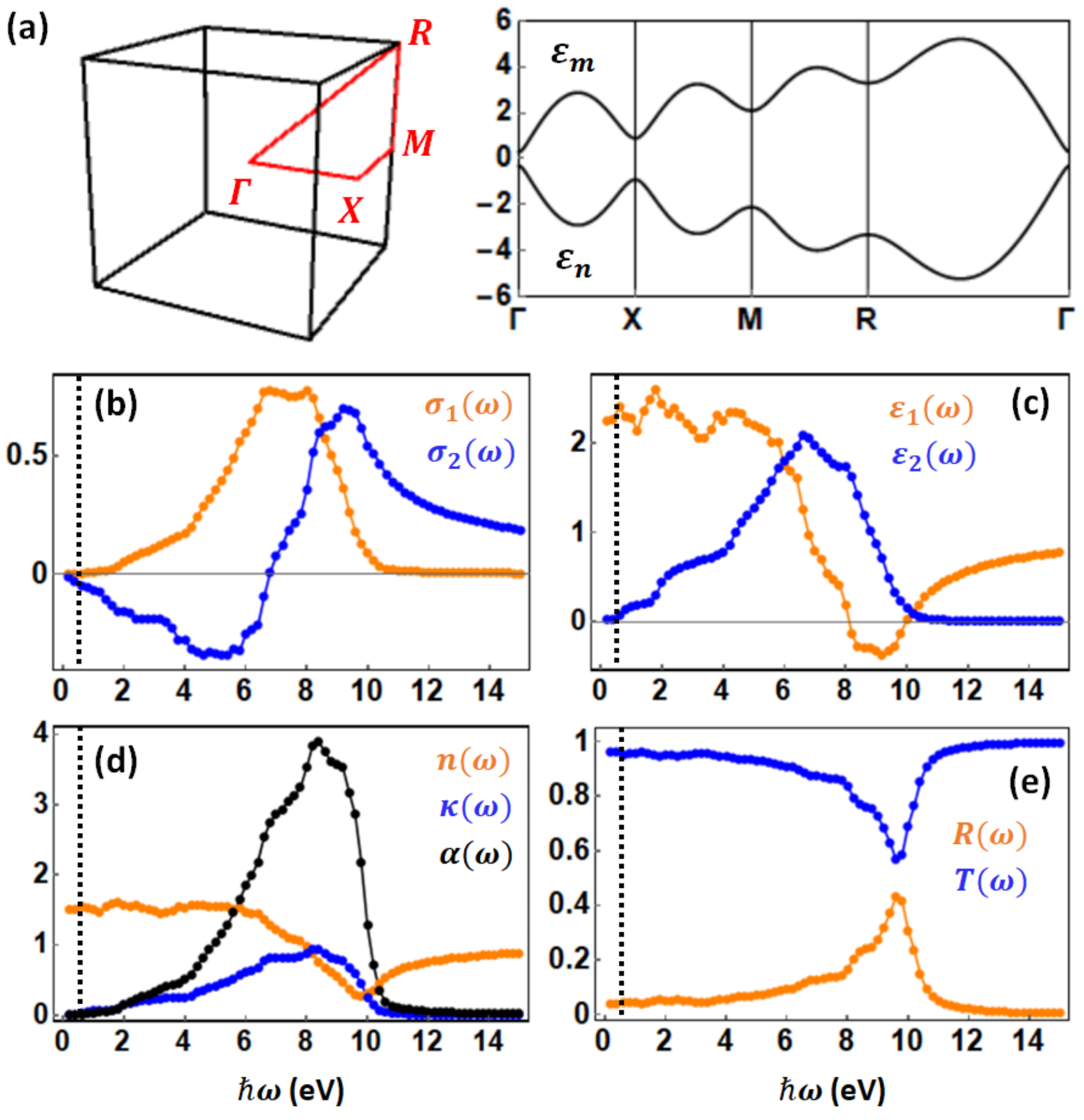}
\caption{(a) The High symmetry points in the cubic BZ and the band structure along a high symmetry line of the minimal lattice model of 3D TI, using parameters similar to that of Bi$_{2}$Te$_{3}$. (b) The optical conductivity $\sigma_{1}(\omega)$ and $\sigma_{2}(\omega)$ of this model in units of $e^{2}/\hbar a$. (c) The dielectric function $\varepsilon_{1}(\omega)$ and $\varepsilon_{2}(\omega)$ assuming $e^{2}/\varepsilon_{0}a$eV $\approx 18$. Note that the relative dielectric constant is given by $\varepsilon_{r}=\lim_{\omega\rightarrow 0}\varepsilon_{1}(\omega)$. (d) The refractive index $n(\omega)$ and $\kappa(\omega)$. (e) The reflectance $T(\omega)$, transmittance at the surface $T(\omega)$, and the absorption coefficient $\alpha(\omega)$ in units of $4eV/\hbar c\approx 2.1\times 10^{7}$m$^{-1}$. The dotted line in each figure indicates the band gap at $0.6$ eV. } 
\label{fig:3DTI_analytic_sigma_epsilon}
\end{center}
\end{figure}

The integrand in Eq.~(\ref{sigma1w_sigma2w_T0}) for this minimal model takes the form\cite{vonGersdorff21_metric_curvature} 
\begin{eqnarray}
&&\sum_{nm}g_{\mu\mu}\delta\left(\omega+\frac{\varepsilon_{n}}{\hbar}-\frac{\varepsilon_{m}}{\hbar}\right)
\nonumber \\
&&=\frac{N}{8d^{2}}\left\{\sum_{i=0}^{3}(\partial_{\mu}d_{i})^{2}
-(\partial_{\mu}d)^{2}\right\}\delta\left(\omega-\frac{2d}{\hbar}\right).\;\;\;\;\;
\nonumber \\
&&\sum_{nm}g_{\mu\mu}^{nm}
\frac{\varepsilon_{m}/\hbar
-\varepsilon_{n}/\hbar}{\omega^{2}-\left(\varepsilon_{m}/\hbar
-\varepsilon_{n}/\hbar\right)^{2}}
\nonumber \\
&&=\frac{N}{8d^{2}}\left\{\sum_{i=0}^{3}(\partial_{\mu}d_{i})^{2}
-(\partial_{\mu}d)^{2}\right\}
\frac{2d/\hbar}{\omega^{2}-4d^{2}/\hbar^{2}}.
\label{gmumu_spec_fn_3D}
\end{eqnarray}
Applying Eq.~(\ref{gmumu_spec_fn_3D}) into Eq.~(\ref{sigma1w_sigma2w_T0}), both $\sigma_{1}(\omega)$ and $\sigma_{2}(\omega)$ can be calculated numerically. For concreteness, we use the parameters $M=0.3$, $A=2.87$, $B=0.3$ in units of eV that are close to Bi$_{2}$Te$_{3}$ as given in Ref.~\onlinecite{Liu10}, and assume a cubic lattice with lattice constant $a=$ nm. The band structure $\left\{\varepsilon_{n},\varepsilon_{m}\right\}$ along a high symmetry line of the Brillouin zone (BZ), together with the numerical results for the frequency dependence of $\left\{\sigma_{1},\sigma_{2},\varepsilon_{1},\varepsilon_{2},\varepsilon_{r},n,\kappa,\alpha,R,T\right\}$, are given in Fig.~\ref{fig:3DTI_analytic_sigma_epsilon}, where the $\delta$-function in Eq.~(\ref{gmumu_spec_fn_3D}) is simulated by a Lorentzian with a small broadening $\eta=0.1$eV.

Several features of $\left\{\sigma_{1},\sigma_{2},\varepsilon_{1},\varepsilon_{2},\varepsilon_{r},n,\kappa,\alpha,R,T\right\}$ shown in Fig.~\ref{fig:3DTI_analytic_sigma_epsilon} (b) to (e) can be qualitatively understood from the band structure shown in Fig.~\ref{fig:3DTI_analytic_sigma_epsilon} (a). Firstly, because the band gap at the $\Gamma$ point is $2M=0.6$eV, and the band width extracted from the band structure is about 10eV, there is no optical absorption in frequency smaller than the band gap $\hbar\omega\leq 0.6$eV or larger than the band width $\hbar\omega\apprge 10$eV. This explains why the real part of the optical conductivity $\sigma_{1}(\omega)$ in Fig.~\ref{fig:3DTI_analytic_sigma_epsilon} (b) and the imaginary part of the dielectric function $\varepsilon_{2}(\omega)=\sigma_{1}(\omega)/\varepsilon_{0}\omega$ in Fig.~\ref{fig:3DTI_analytic_sigma_epsilon} (c) are finite only within the range $0.6$eV$\leq\hbar\omega\apprle 10$eV. However, it should be reminded that this argument does not apply to the imaginary part of the optical conductivity $\sigma_{2}(\omega)$ shown in Fig.~\ref{fig:3DTI_analytic_sigma_epsilon} (b). As can be understood from Eq.~(\ref{gmumu_spec_fn_3D}), $\sigma_{2}(\omega)$ is finite at any frequency, and so is the real part of the dielectric function\cite{Yu10} $\varepsilon_{1}(\omega)=1-\sigma_{2}(\omega)/\varepsilon_{0}\omega$. In fact, it is obvious that $\varepsilon_{1}(\omega)$ is finite inside the gap, since its zero frequency limit is the relative dielectric constant $\varepsilon_{r}$ that gives the capacitance of the TI. Finally, because $\left\{n,\kappa,\alpha,R,T\right\}$ depend on both $\sigma_{1}(\omega)$ and $\sigma_{2}(\omega)$ according to the formulas in Sec.~\ref{sec:optical_properties}, they are in general finite at any frequency as shown in Fig.~\ref{fig:3DTI_analytic_sigma_epsilon} (d) and (e).

Despite serving our purpose of demonstrating the quantum geometric origin of the dielectric and optical properties, these numerical results are at odds with the realistic experimental data measured in prototype TIs\cite{Greenaway65,Ou14,Zhao15}. The dielectric function for Bi$_{2}$Te$_{3}$, Sb$_{2}$Te$_{3}$, and Bi$_{1.5}$Sb$_{0.5}$Te$_{1.8}$Se$_{1.2}$ (see Ref.~\onlinecite{Yin17} for a summary) show that the $\epsilon_{1}$ peaks at around $\hbar\omega\sim 1$eV with a magnitude about 30, and $\epsilon_{2}$ peaks at about $\hbar\omega\sim 1.5$eV with a magnitude about 40. In comparison, our numerical results for the dielectric function is about 20 times smaller than these experimental results. The relative dielectric constant obtained from our calculation is about $\varepsilon_{r}\approx 2.2$, which is also much smaller than that reported experimentally\cite{Greenaway65} $\varepsilon_{r}\approx 100$. As a result, one may proceed to integrate the imaginary part of dielectric function $\varepsilon_{2}(\omega)$ to obtain the spread of valence-band Wannier function $\Omega_{I}$ according to Eq.~(\ref{OmegaI_wintegral_epsilon2}), but the result will be much smaller than that extracted from the experimental data\cite{CardenasCastillo24_spread_Wannier}. We suspect that these discrepancies are due to various details that cannot be captured by our low energy minimal model, such as other nontopological orbitals and the quintuple layer structure, and other correlation effects that can affect the quantum metric\cite{Chen22_dressed_Berry_metric}. Most importantly, the metallic surface states must play an important role in the ellipsometry measurement of optical effects which is a surface measurement, such as generating surface plasmonic modes\cite{Yin17}. Nevertheless, we emphasize that this simple low-energy model serves our purpose to demonstrate the dielectric and optical properties obtained entirely from $\left\{\sigma_{1},\sigma_{2}\right\}$, and therefore they are completely caused by the interband transition determined by the valence band quantum metric. Finally, note that Fig.~\ref{fig:3DTI_analytic_sigma_epsilon} shows that these dielectric and optical properties peak at frequency $\hbar\omega\sim 6$ to 10 eV, which is much larger than the band gap 0.6 eV, suggesting that the band inversion at low energy that causes the topological order is not the main contribution to these properties. As a result, we also do not expect that these properties can be used to distinguish the topological phases of matter. Finally, as a comparison and for pedagogical reasons, in Appendix \ref{apx:linear_Dirac_sigma_epsilon} we present the calculation of linear Dirac models assuming the linear Dirac cone extends all the way to the BZ boundary, in which analytical expressions of optical conductivity can be given. The result indicates a reduced optical transition matrix element in the linear Dirac model, i.e., the deformation of Dirac cone by hopping terms in the lattice model is actually crucial to determine the magnitude of the dielectric and optical properties.

\subsection{Dielectric and optical markers for 3D TI}

Despite the inability to capture the dielectric functions of real TIs, we shall proceed to use the minimal model for 3D class AII to demonstrate the feasibility of dielectric and optical markers, since it is the simplest model that can serve our purpose. The regularization of the momentum space model in Eq.~(\ref{3D_class_AII_kspace_model}) on a cubic lattice yields\cite{Chen20_absence_edge_current}
\begin{eqnarray}
&&H=\sum_{i,\sigma}\tilde{M}\left\{c_{is\sigma}^{\dag}c_{is\sigma}-c_{ip\sigma}^{\dag}c_{ip\sigma}\right\}
\nonumber \\
&&+\sum_{i,I}t_{\parallel}\left\{c_{iI\uparrow}^{\dag}c_{i+a\overline{I}\downarrow}
-c_{i+aI\uparrow}^{\dag}c_{i\overline{I}\downarrow}+h.c.\right\}
\nonumber \\
&&+\sum_{i,I}t_{\parallel}\left\{-ic_{iI\uparrow}^{\dag}c_{i+b\overline{I}\downarrow}
+ic_{i+bI\uparrow}^{\dag}c_{i\overline{I}\downarrow}+h.c.\right\}
\nonumber \\
&&+\sum_{i,\sigma}t_{\perp}\left\{-c_{is\sigma}^{\dag}c_{i+cp\sigma}+c_{i+cs\sigma}^{\dag}c_{ip\sigma}+h.c.\right\}
\nonumber \\
&&-\sum_{i,\sigma}M_{1}\left\{c_{is\sigma}^{\dag}c_{i+cs\sigma}-c_{ip\sigma}^{\dag}c_{i+cp\sigma}+h.c.\right\}
\nonumber \\
&&-\sum_{i,\delta,\sigma}M_{2}\left\{c_{is\sigma}^{\dag}c_{i+\delta s\sigma}-c_{ip\sigma}^{\dag}c_{i+\delta p\sigma}+h.c.\right\},
\label{3DTIFMM_Hamiltonian}
\end{eqnarray}
where $\tilde{M}=M+6B$, $M_{1}=M_{2}=B$, $t_{\parallel}=t_{\perp}=A/2$, $I=\left\{s,p\right\}$ and $\overline{I}=\left\{p,s\right\}$ are the orbital indices, $\delta=\left\{a,b,c\right\}$ denotes the lattice constants along the three crystalline directions, and $\sigma=\left\{\uparrow,\downarrow\right\}$ is the spin index. We choose the parameters of larger band gap $t_{\parallel}=t_{\perp}=M_{1}=M_{2}=-M=1$ assuming in units of eV, such that reliable results can be obtained from a smaller cubic lattice of size $5\times 5\times 5$ with little numerical effort. Periodic boundary conditions are imposed in all three directions, so no topological surface state appears. In particular, we consider a single impurity with an impurity potential that is the same for all four degrees of freedom on the impurity site
\begin{eqnarray}
H_{imp}=\sum_{\sigma}U_{imp}c_{iI\sigma}^{\dag}c_{iI\sigma},
\label{3DTI_Himp}
\end{eqnarray}
with $U_{imp}=8$. We will use the markers to investigate how the dielectric and optical properties are changed near the impurity site.

%Ideally, the markers should be independent of ${\bf r}$ in a homogeneous system in the thermodynamic limit, and should be equal to the values obtained from momentum integration in Sec.~\ref{sec:3DTI_kspace}, but this is no longer true in the presence of impurities. 

\begin{figure*}[ht]
\begin{center}
\includegraphics[clip=true,width=1.6\columnwidth]{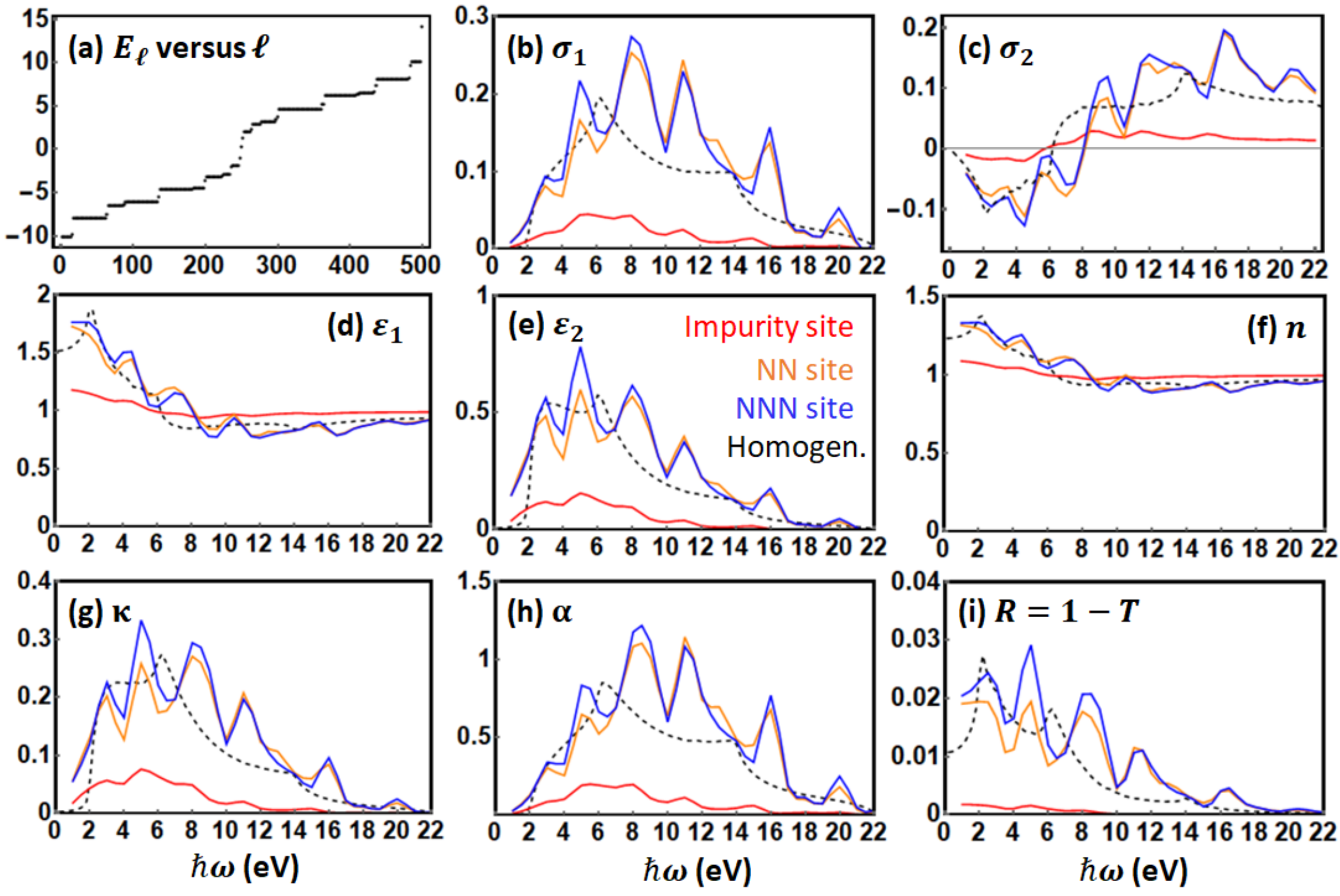}
\caption{Dielectric and optical markers for the minimal model of 3D TI near the strong impurity $U_{imp}=8$ simulated on a $5\times 5\times 5$ lattice, where we plot the frequency dependence on the impurity site (red), nearest-neighbor site of the impurity (orange), and next-nearest-neighbor site of the impurity (blue), as well as the homogeneous values calculated from momentum integration (dashed black). (a) The eigenenergies sorted from small to large, where the stairwise feature indicates a large degeneracy of eigenstates due to crystalline symmetry, causing the peaks in the frequency dependence of the markers. (b) The real part $\sigma_{1}({\bf r},\omega)$ and (c) imaginary part $\sigma_{2}({\bf r},\omega)$ of optical conductivity marker in units of $e^{2}/\hbar a$. (d) The real part $\varepsilon_{1}({\bf r},\omega)$ and (e) imaginary part $\varepsilon_{2}({\bf r},\omega)$ of dielectric function marker assuming $e^{2}/\varepsilon_{0}at_{\parallel}\approx 18$. The refractive index markers (f) $n({\bf r},\omega)$ and (g) $\kappa({\bf r},\omega)$. (h) The absorption coefficient marker $\alpha({\bf r},\omega)$ in units of $e^{2}/\varepsilon_{0}c\hbar a\approx 10^{8}$m$^{-1}$. (i) The reflectance marker $R({\bf r},\omega)=1-T({\bf r},\omega)$.} 
\label{fig:optical_markers_TI_figure}
\end{center}
\end{figure*}

Cautions must be taken when applying the markers in Eq.~(\ref{optical_conductivity_markers}) to such a small $5\times 5\times 5$ lattice. Firstly, in Fig.~\ref{fig:optical_markers_TI_figure} (a) we show the $5\times 5\times 5\times 4=500$ lattice eigenenergies $E_{\ell}$ sorted from small to large, where the stairwise feature is a clear indication of large degeneracy of the eigenstates. This degeneracy comes from symmetry operations on the cubic lattice with respect to the impurity site, such as inversion and rotation, as well as spin degeneracy. As a result, the optical transition will peak at discrete frequencies equal to the energy difference between the stairs in Fig.~\ref{fig:optical_markers_TI_figure} (a), which explains the peaks in the real part of optical conductivity marker $\sigma_{1}({\bf r},\omega)$ shown in Fig.~\ref{fig:optical_markers_TI_figure} (b). To obtain a smooth frequency-dependence, one can either employ a larger lattice that is numerically more costly, or choose an artificial broadening $\eta$ in the $\delta$-function of Eq.~(\ref{optical_conductivity_markers}) that is comparable to the energy spacing between the stairs, leading us to choose $\eta=0.5$. With this choice of $\eta$, we obtain an absolute scale of $\sigma_{1}({\bf r},\omega)$ comparable to its homogeneous value calculated from the momentum integration formula in Sec.~\ref{sec:3DTI_kspace}, indicating the feasibility of the marker. To further quantify the spatial dependence, we plot the markers on the impurity site ${\bf r}=(0,0,0)$, the nearest-neighbor site from the impurity ${\bf r}=(a,0,0)$, and the next-nearest-neighbor site ${\bf r}=(a,a,0)$, as shown by the red, orange, and blue lines in Fig.~\ref{fig:optical_markers_TI_figure}, respectively.

The degenerate eigenstates pose a even more serious problem to the imaginary part of optical conductivity marker $\sigma_{2}({\bf r},\omega)$, since from Eq.~(\ref{optical_conductivity_markers}) one sees that the energy-dependent kernel $\left[E_{\ell '}/\hbar
-E_{\ell}/\hbar\right]/\left[\omega^{2}-\left(E_{\ell '}/\hbar
-E_{\ell}/\hbar\right)^{2}\right]$ has no artificial broadening to mitigate. Thus a straightforward calculation of Eq.~(\ref{optical_conductivity_markers}) on a small lattice will yield $\sigma_{2}({\bf r},\omega)$ that fluctuates wildly in frequency, which is not of much usage. In other words, the formula for $\sigma_{2}({\bf r},\omega)$ in Eq.~(\ref{optical_conductivity_markers}) is only useful when lattice size is large enough. To overcome this numerical artifact, we propose the following three-step trick to obtain a reliable $\sigma_{2}({\bf r},\omega)$.

(i) For any position ${\bf r}$ under question, first calculate the real part $\sigma_{1}({\bf r},\omega)$ on discrete frequency points $\omega$ using appropriate artificial broadening as discussed above, over the full frequency range that $\sigma_{1}({\bf r},\omega)$ is nonzero. 

(ii) Interpolate the $\sigma_{1}({\bf r},\omega)$ on discrete frequency points into a continuous function of frequency $\overline{\sigma}_{1}({\bf r},\omega)$. This can be easily done using any numerical package of interpolation. 

(iii) Use Kramers-Kronig relation to calculate $\sigma_{2}({\bf r},\omega)$ from the frequency integration (this is simply the ${\bf r}$-dependent version of Eq.~(\ref{Kramers_Kronig_sigma}))
\begin{eqnarray}
\sigma_{2}({\bf r},\omega)=-\frac{2\omega}{\pi}{\cal P}\int_{0}^{\infty}d\omega '\frac{\overline{\sigma}_{1}({\bf r},\omega ')}{\omega^{\prime 2}-\omega^{2}}.
\end{eqnarray}

Using this numerical recipe, we are able to obtain a $\sigma_{2}({\bf r},\omega)$ consistent with its homogeneous value calculated using momentum integration, as shown in Fig.~\ref{fig:optical_markers_TI_figure} (c). Once both $\sigma_{1}({\bf r},\omega)$ and $\sigma_{2}({\bf r},\omega)$ are calculated reliably, we can use them to get other markers $\left\{\varepsilon_{1},\varepsilon_{2},\varepsilon_{r},n,\kappa,\alpha,R,T\right\}$ on any site ${\bf r}$ and frequency $\omega$ according to the formula in Sec.~\ref{sec:optical_properties}, as given in Fig.~\ref{fig:optical_markers_TI_figure} (d) to (i).

Our results for the markers shown in Fig.~\ref{fig:optical_markers_TI_figure} indicate two surprisingly universal features caused by the potential impurity. The first is that the dielectric and optical responses are locally suppressed by the impurity in almost all the frequencies, since the magnitude of the markers on the impurity site (red lines) is smaller than that on the neighboring sites (orange and blue lines) in most frequencies. The second is that despite this suppression, the dielectric and optical responses quickly recovers their homogeneous values once the position ${\bf r}$ moves away from the impurity site (orange and blues lines are similar to the black dash lines). These features suggest that impurity effects in TIs are genuinely local, which seem to be fairly intuitive since the wave functions in insulators are generally more local, e.g., no long range Friedel oscillation occurs. These features thus indicate a number of impurity effects on TIs that can be described in simple terms. For instance, a reduced reflectance $R$ (enhanced transmittance $T$) and a lessen absorption coefficient $\alpha$ (increased decay length) mean the impurities are making the TI more transparent to any color of light. A reduced dielectric constant $\varepsilon_{r}=\sum_{\bf r}\varepsilon_{1}({\bf r},\omega=0)/N$ means the capacitance is deteriorated by the impurities. Finally, according to Eq.~(\ref{OmegaI_wintegral_epsilon2_atr}), a reduced $\varepsilon_{2}({\bf r},\omega)$ implies that the valence band Wannier functions become more localized because of the impurities, at least within this minimal model.

\section{Conclusions}

In summary, we elaborate the quantum geometric origin of many dielectric and optical properties of 3D semiconductors and insulators, including the dielectric function $\left\{\varepsilon_{1},\varepsilon_{2}\right\}$, charge susceptibility $\left\{\chi_{1},\chi_{2}\right\}$, refractive index $\left\{n,\kappa\right\}$, absorption coefficient $\alpha$, relative dielectric constant $\varepsilon_{r}$, reflectance $R$, and transmittance $T$. Our statement is made because the quantum metric of the fully antisymmetric valence band state can be decomposed into the matrix elements of optical transition between any pair of valence and conduction bands, and these matrix elements enter both real and imaginary parts of optical conductivity. Since these non-excitonic dielectric and optical properties can be expressed in terms of the complex optical conductivity, their quantum geometric origin is evident. Thus quantum geometry even influences the optical properties in the macroscopic scale that can be perceived by human eyes, such as the refractive index and reflectance.

Our formalism further introduces the dielectric and optical markers defined on lattice sites. These markers quantifies the propagation of electromagnetic waves in the nanometer scale inside a disordered material, including both the frequency and spatial dependence of the propagation, as demonstrated by a minimal model of 3D TIs. Through calculating the real and imaginary parts of the optical conductivity marker $\sigma_{1}({\bf r},\omega)$ and $\sigma_{2}({\bf r},\omega)$, we show that all other dielectric and optical markers can be derived following their relations with the former, and they can explain the spatial inhomogeneity revealed by NSOM. Furthermore, the frequency-integration of the dielectric function marker gives the fidelity marker $\sum_{\mu}{\cal G}_{\mu\mu}({\bf r})$ proposed previously, as well as the generalization of the spread of valence-band Wannier function $\Omega_{I}({\bf r})$ to disordered insulators. In comparison with usual classical approaches, these markers have the advantage of capturing the quantum effects caused by the disorder, and they reveal a local suppression of dielectric and optical properties by impurities in the minimal model of TIs. On the other hand, our minimal model also shows its own limitation, indicating that shall one intend to capture accurately all these dielectric and optical properties to compare with experiment, one must first accurately describe the complete band structure and the wave functions of all the bands, not just a subset of bands in the low energy sector. We anticipate that these dielectric and optical markers will have a very broad applications to study countless semiconductors and insulators under the influence of various kinds of impurities, which await to be further explored.

%Our minimal model of 3D TI also demonstrates how impurities can help to engineer the dielectric and optical properties, such as changing the capacitance and color of the material, that we anticipate would have a significant technological impact. 

\appendix

\section{Details of numerical calculation of optical conductivity markers \label{apx:numerical_optical_conductivity_marker}}

The optical conductivity markers in Eq.~(\ref{optical_conductivity_markers}) and the nonlocal optical conductivity in Eq.~(\ref{nonlocal_optical_conductivity}) can be implemented conveniently in numerical calculations by the following trick\cite{deSousa23_fidelity_marker}. We first define the operators
\begin{widetext}
\begin{eqnarray}
&&{\cal M}_{1\mu}(\omega)=\sum_{\ell<\ell '}S_{\ell}{\hat\mu}S_{\ell '}\sqrt{[f(E_{\ell})-f(E_{\ell '})]\delta\left(\omega+\frac{E_{\ell}}{\hbar}
-\frac{E_{\ell '}}{\hbar}\right)},
\nonumber \\
&&{\cal M}_{2\mu}(\omega)=\sum_{\ell<\ell '}S_{\ell}{\hat\mu}S_{\ell '}\sqrt{\left|\left[f(E_{\ell})-f(E_{\ell '})\right]\frac{E_{\ell '}/\hbar
-E_{\ell}/\hbar}{\omega^{2}-\left(E_{\ell '}/\hbar
-E_{\ell}/\hbar\right)^{2}}\right|},
\nonumber \\
&&\overline{\cal M}_{2\mu}(\omega)=\sum_{\ell<\ell '}S_{\ell '}{\hat\mu}S_{\ell}\sqrt{\left|\left[f(E_{\ell})-f(E_{\ell '})\right]\frac{E_{\ell '}/\hbar
-E_{\ell}/\hbar}{\omega^{2}-\left(E_{\ell '}/\hbar
-E_{\ell}/\hbar\right)^{2}}\right|}\times{\rm Sgn}\left\{\left[f(E_{\ell})-f(E_{\ell '})\right]\frac{E_{\ell '}/\hbar
-E_{\ell}/\hbar}{\omega^{2}-\left(E_{\ell '}/\hbar
-E_{\ell}/\hbar\right)^{2}}\right\},
\nonumber \\
\end{eqnarray}
according to which the optical conductivity markers become 
\begin{eqnarray}
&&\sigma_{1}({\bf r},\omega)=\left(\frac{\pi e^{2}}{\hbar V_{\rm cell}}\right)\omega\sum_{\gamma}\langle{\bf r},\gamma|{\cal M}_{1\mu}(\omega){\cal M}_{1\mu}^{\dag}(\omega)|{\bf r},\gamma\rangle,
\nonumber \\
&&\sigma_{2}({\bf r},\omega)=\left(\frac{2e^{2}}{\hbar V_{\rm cell}}\right)\omega\sum_{\gamma}\langle{\bf r},\gamma|{\cal M}_{2\mu}(\omega)\overline{\cal M}_{2\mu}(\omega)|{\bf r},\gamma\rangle.
\end{eqnarray}
Likewisely, for the nonlocal optical conductivity, we further define the operators
\begin{eqnarray}
&&{\cal M}_{1\mu,{\bf r'}}^{\dag}(\omega)=\sum_{\ell<\ell '}S_{\ell '}{\hat\mu}_{\bf r'}S_{\ell}\sqrt{[f(E_{\ell})-f(E_{\ell '})]\delta\left(\omega+\frac{E_{\ell}}{\hbar}
-\frac{E_{\ell '}}{\hbar}\right)},
\nonumber \\
&&\overline{\cal M}_{2\mu,{\bf r'}}(\omega)=\sum_{\ell<\ell '}S_{\ell '}{\hat\mu}_{\bf r'}S_{\ell}\sqrt{\left|\left[f(E_{\ell})-f(E_{\ell '})\right]\frac{E_{\ell '}/\hbar
-E_{\ell}/\hbar}{\omega^{2}-\left(E_{\ell '}/\hbar
-E_{\ell}/\hbar\right)^{2}}\right|}\times{\rm Sgn}\left\{\left[f(E_{\ell})-f(E_{\ell '})\right]\frac{E_{\ell '}/\hbar
-E_{\ell}/\hbar}{\omega^{2}-\left(E_{\ell '}/\hbar
-E_{\ell}/\hbar\right)^{2}}\right\}.
\nonumber \\
\end{eqnarray}
Once again the ${\hat\mu}_{\bf r'}$ is the position operator at site ${\bf r'}$. Using these operators, the nonlocal optical conductivity is
\begin{eqnarray}
&&\sigma_{1}({\bf r,r'},\omega)=\left(\frac{\pi e^{2}}{\hbar V_{\rm cell}}\right)\omega\sum_{\gamma}\langle{\bf r},\gamma|{\cal M}_{1\mu}(\omega){\cal M}_{1\mu,{\bf r'}}^{\dag}(\omega)|{\bf r},\gamma\rangle,
\nonumber \\
&&\sigma_{2}({\bf r,r'},\omega)=\left(\frac{2e^{2}}{\hbar V_{\rm cell}}\right)\omega\sum_{\gamma}\langle{\bf r},\gamma|{\cal M}_{2\mu}(\omega)\overline{\cal M}_{2\mu,{\bf r'}}(\omega)|{\bf r},\gamma\rangle.
\end{eqnarray}
\end{widetext}
We will focus only on the calculation of the optical conductivity markers $\sigma_{i}({\bf r},\omega)$ in the present work, while leaving the nonlocal $\sigma_{i}({\bf r,r'},\omega)$ for future investigations.

\section{Dielectric and optical properties of linear Dirac models \label{apx:linear_Dirac_sigma_epsilon}}

The linear Dirac models have been a paradigm to describe the low energy sector of topological materials\cite{Schnyder08,Ryu10,Chiu16}, from which analytical expressions of many material properties can be derived. In this section, we use linear Dirac model to investigate the dielectric and optical properties of 3D TIs at zero temperature $T=0$, and make comparison with the more accurate lattice model results shown in Fig.~\ref{fig:3DTI_analytic_sigma_epsilon}. The linear Dirac model corresponds to simply replacing the ${\bf d}$-vector in Eq.~(\ref{3D_class_AII_kspace_model}) by
\begin{eqnarray}
d_{0}=M,\;\;\;d_{1}=vk_{x},\;\;\;d_{2}=vk_{y},\;\;\;d_{3}=vk_{z},
\end{eqnarray}
where $v$ is the Fermi velocity of the Dirac cone. The fidelity number spectral function ${\cal G}_{\mu\mu}^{d}(\omega)$ in Eq.~(\ref{sigma1w_sigma2w}) of this model has been calculated in Ref.~\onlinecite{deSousa23_fidelity_marker}, so we can simply use it to obtain the real part of the optical conductivity $\sigma_{1}(\omega)$. In addition, to make sense of the calculation, it is necessary to impose a cut-off frequency $\omega_{f}$ beyond which $\sigma_{1}(\omega>\omega_{f})=0$ to simulate the finite band width caused by the BZ boundary, for otherwise optical absorption can occur in arbitrarily large frequency. These concerns lead to 
\begin{widetext}
\begin{eqnarray}
\sigma_{1}(\omega)=\frac{Ne^{2}\omega}{12\pi^{2}\hbar v}\left\{\frac{3\pi M^{2}\sqrt{\hbar^{2}\omega^{2}-4M^{2}}}{2\hbar^{3}\omega^{3}}
+\frac{\pi(\hbar^{2}\omega^{2}-4M^{2})^{3/2}}{4\hbar^{3}\omega^{3}}\right\}_{2|M|/\hbar\leq\omega\leq\omega_{f}}.
\label{sigma1_linear_Dirac}
\end{eqnarray}
The imaginary part of the optical conductivity $\sigma_{2}(\omega)$ can then be obtained from the Kramers-Kronig relation in Eq.~(\ref{Kramers_Kronig_sigma}) with the frequency integration containing the cut-off $\int_{2|M|/\hbar}^{\omega f}d\omega '$, since $\sigma_{1}(\omega ')$ is finite only within this frequency range, yielding the result for the frequency higher than the band gap $\hbar\omega\geq 2|M|$
\begin{eqnarray}
\sigma_{2}(\omega)&=&\frac{Ne^{2}\omega}{12\pi^{2}\hbar v}\left\{-\frac{M^{2}\sqrt{\hbar^{2}\omega_{f}^{2}-4M^{2}}}{2\hbar^{3}\omega_{f}\omega^{2}}
-\frac{1}{2}\ln\left(\frac{\hbar\omega_{f}}{2|M|}+\sqrt{\frac{\hbar^{2}\omega_{f}^{2}}{4M^{2}}-1}\right)\right.
\nonumber \\
&&\left.+\frac{M^{2}\sqrt{\hbar^{2}\omega^{2}-4M^{2}}}{\hbar^{3}\omega^{3}}\left(\frac{\hbar^{2}\omega^{2}}{4M^{2}}+\frac{1}{2}\right)
\ln\left|\frac{\hbar\omega\sqrt{\hbar^{2}\omega_{f}^{2}-4M^{2}}+\hbar\omega_{f}\sqrt{\hbar^{2}\omega^{2}-4M^{2}}}
{\hbar\omega\sqrt{\hbar^{2}\omega_{f}^{2}-4M^{2}}-\hbar\omega_{f}\sqrt{\hbar^{2}\omega^{2}-4M^{2}}}\right|\right\}_{\hbar\omega\geq 2|M|},
\label{sigma2_linear_Dirac_wlarge}
\end{eqnarray}
and the result for the frequency smaller than the band gap $\hbar\omega\leq 2|M|$
\begin{eqnarray}
\sigma_{2}(\omega)&=&\frac{Ne^{2}\omega}{12\pi^{2}\hbar v}\left\{-\frac{M^{2}\sqrt{\hbar^{2}\omega_{f}^{2}-4M^{2}}}{2\hbar^{3}\omega_{f}\omega^{2}}
-\frac{1}{2}\ln\left(\frac{\hbar\omega_{f}}{2|M|}+\sqrt{\frac{\hbar^{2}\omega_{f}^{2}}{4M^{2}}-1}\right)\right.
\nonumber \\
&&\left.-\frac{M^{2}\sqrt{4M^{2}-\hbar^{2}\omega^{2}}}{\hbar^{3}\omega^{3}}\left(2\cdot\frac{\hbar^{2}\omega^{2}}{4M^{2}}+1\right)
\left[\tan^{-1}\left(\frac{\hbar\omega_{f}\sqrt{4M^{2}-\hbar^{2}\omega^{2}}}{\hbar\omega\sqrt{\hbar^{2}\omega_{f}^{2}-4M^{2}}}\right)-\frac{\pi}{2}\right]\right\}_{\hbar\omega\leq 2|M|}.
\label{sigma2_linear_Dirac_wsmall}
\end{eqnarray}
\end{widetext}
To fix the absolute scale of the optical conductivity, we demand that the linear Dirac cone extended to the boundary of the BZ $k_{x}=\hbar\pi/a$ should give a band width $\pi\hbar v/a\approx \hbar\omega_{f}/2\approx 5$eV according to the the band structure of the lattice model shown in Fig.~\ref{fig:3DTI_analytic_sigma_epsilon} (a). This fixes the values of Fermi velocity divided by lattice constant $v/a$ and the cut-off frequency $\hbar\omega_{f}\approx 10$eV, and then the $\sigma_{i}(\omega)$ can be plotted in the same units $e^{2}/\hbar a$ as Fig.~\ref{fig:3DTI_analytic_sigma_epsilon} (b).

\begin{figure}[ht]
\begin{center}
\includegraphics[clip=true,width=0.99\columnwidth]{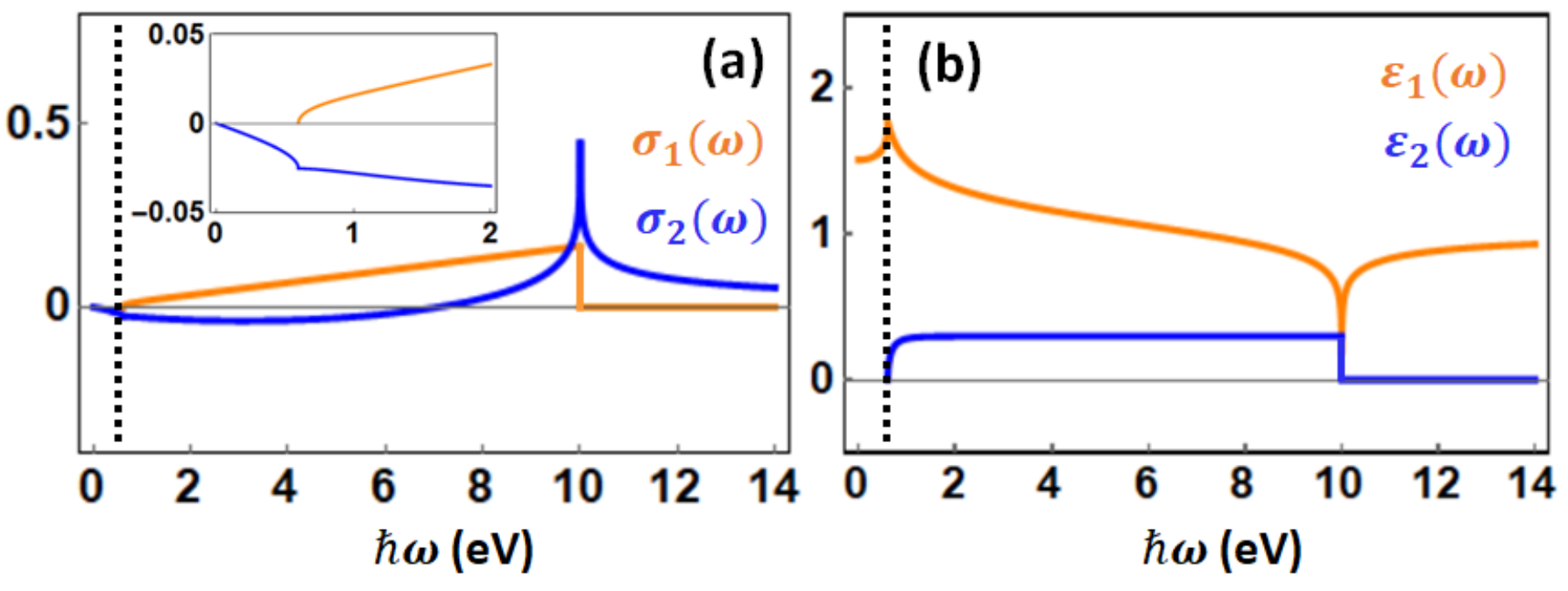}
\caption{Dielectric and optical properties of linear Dirac model at zero temperature, assuming a band width at the BZ boundary $\pi\hbar v/a\approx 5$eV that sets a cut-off frequency $\hbar\omega_{f}\approx$10eV to optical absorption, plotted in the same scale as Fig.~\ref{fig:3DTI_analytic_sigma_epsilon}. (a) The optical conductivity $\sigma_{1}(\omega)$ and $\sigma_{2}(\omega)$ (inset shows the result at low frequency), and (b) the dielectric functions $\varepsilon_{1}(\omega)$ and $\varepsilon_{2}(\omega)$.} 
\label{fig:linear_Dirac_sigma12}
\end{center}
\end{figure}

The results shown in Fig.~\ref{fig:linear_Dirac_sigma12} (a) indicate that the real $\sigma_{1}(\omega)$ and imaginary $\sigma_{2}(\omega)$ parts of the optical conductivity have a similar frequency-dependence but a much smaller magnitude (about 20$\%$) compared to those of the lattice model shown in Fig.~\ref{fig:3DTI_analytic_sigma_epsilon} (b). This suggests that the hopping terms of the lattice model that causes the deformation of Dirac cone at large momentum has a significant influence on the wave function and optical absorption matrix element, i.e., the quantum metric, in real materials. It is also intriguing to notice that $\sigma_{1}(\omega)$ at large frequency $\hbar\omega\gg 2|M|$ is roughly linear to frequency, a behavior that resembles the $\sigma_{1}(\omega)$ of Dirac and Weyl semimetals\cite{Hosur12,Bacsi13,Timusk13,Ashby14,Xu16,deSousa23_graphene_opacity} that can be simulated by a zero mass term $M=0$, which is reasonable since at large frequency the Dirac cone is fairy linear and the mass term $M$ would become unimportant. Comparing the dielectric functions shown in Fig.~\ref{fig:linear_Dirac_sigma12} (b) with that of the lattice model shown in Fig.~\ref{fig:3DTI_analytic_sigma_epsilon} (c), we see that the real part $\varepsilon_{1}(\omega)$ has a comparable magnitude, but the imaginary part $\varepsilon_{2}(\omega)$ is much smaller, again due to the reduced optical transition matrix element. We remark that these properties show no difference in the topologically trivial $M>0$ and nontrivial $M<0$ phases, since Eqs.~(\ref{sigma1_linear_Dirac}), (\ref{sigma2_linear_Dirac_wlarge}), and (\ref{sigma2_linear_Dirac_wsmall}) depend on $M^{2}$ or $|M|$, so we do not expect that bulk dielectric and optical properties can distinguish the topological phases of 3D TIs.

%To make sense of the calculation, it is necessary to impose a BZ boundary such that the band width is finite, for otherwise optical absorption can occur in arbitrarily large frequency. Thus we assume a spherical BZ boundary at $k=\hbar\pi/a$ that confines the momentum integration of Eq.~(\ref{sigma1w_sigma2w_T0}) in the spherical coordinates to
%\begin{eqnarray}
%\int d^{3}{\bf k}=\int_{0}^{2\pi}d\phi\int_{0}^{\pi}d\theta\sin\theta\int_{0}^{\hbar\pi/a}dk\,k^{2}.
%\end{eqnarray}

\bibliography{Literatur_abbreviated}

\begin{thebibliography}{51}%
\makeatletter
\providecommand \@ifxundefined [1]{%
 \@ifx{#1\undefined}
}%
\providecommand \@ifnum [1]{%
 \ifnum #1\expandafter \@firstoftwo
 \else \expandafter \@secondoftwo
 \fi
}%
\providecommand \@ifx [1]{%
 \ifx #1\expandafter \@firstoftwo
 \else \expandafter \@secondoftwo
 \fi
}%
\providecommand \natexlab [1]{#1}%
\providecommand \enquote  [1]{``#1''}%
\providecommand \bibnamefont  [1]{#1}%
\providecommand \bibfnamefont [1]{#1}%
\providecommand \citenamefont [1]{#1}%
\providecommand \href@noop [0]{\@secondoftwo}%
\providecommand \href [0]{\begingroup \@sanitize@url \@href}%
\providecommand \@href[1]{\@@startlink{#1}\@@href}%
\providecommand \@@href[1]{\endgroup#1\@@endlink}%
\providecommand \@sanitize@url [0]{\catcode `\\12\catcode `\$12\catcode
  `\&12\catcode `\#12\catcode `\^12\catcode `\_12\catcode `\%12\relax}%
\providecommand \@@startlink[1]{}%
\providecommand \@@endlink[0]{}%
\providecommand \url  [0]{\begingroup\@sanitize@url \@url }%
\providecommand \@url [1]{\endgroup\@href {#1}{\urlprefix }}%
\providecommand \urlprefix  [0]{URL }%
\providecommand \Eprint [0]{\href }%
\providecommand \doibase [0]{http://dx.doi.org/}%
\providecommand \selectlanguage [0]{\@gobble}%
\providecommand \bibinfo  [0]{\@secondoftwo}%
\providecommand \bibfield  [0]{\@secondoftwo}%
\providecommand \translation [1]{[#1]}%
\providecommand \BibitemOpen [0]{}%
\providecommand \bibitemStop [0]{}%
\providecommand \bibitemNoStop [0]{.\EOS\space}%
\providecommand \EOS [0]{\spacefactor3000\relax}%
\providecommand \BibitemShut  [1]{\csname bibitem#1\endcsname}%
\let\auto@bib@innerbib\@empty
%</preamble>
\bibitem [{\citenamefont {Provost}\ and\ \citenamefont
  {Vallee}(1980)}]{Provost80}%
  \BibitemOpen
  \bibfield  {author} {\bibinfo {author} {\bibfnamefont {J.~P.}\ \bibnamefont
  {Provost}}\ and\ \bibinfo {author} {\bibfnamefont {G.}~\bibnamefont
  {Vallee}},\ }\bibfield  {title} {\enquote {\bibinfo {title} {Riemannian
  structure on manifolds of quantum states},}\ }\href
  {https://projecteuclid.org:443/euclid.cmp/1103908308} {\bibfield  {journal}
  {\bibinfo  {journal} {Comm. Math. Phys.}\ }\textbf {\bibinfo {volume} {76}},\
  \bibinfo {pages} {289--301} (\bibinfo {year} {1980})}\BibitemShut {NoStop}%
\bibitem [{\citenamefont {Ozawa}\ and\ \citenamefont
  {Goldman}(2018)}]{Ozawa18}%
  \BibitemOpen
  \bibfield  {author} {\bibinfo {author} {\bibfnamefont {T.}~\bibnamefont
  {Ozawa}}\ and\ \bibinfo {author} {\bibfnamefont {N.}~\bibnamefont
  {Goldman}},\ }\bibfield  {title} {\enquote {\bibinfo {title} {Extracting the
  quantum metric tensor through periodic driving},}\ }\href {\doibase
  10.1103/PhysRevB.97.201117} {\bibfield  {journal} {\bibinfo  {journal} {Phys.
  Rev. B}\ }\textbf {\bibinfo {volume} {97}},\ \bibinfo {pages} {201117}
  (\bibinfo {year} {2018})}\BibitemShut {NoStop}%
\bibitem [{\citenamefont {von Gersdorff}\ and\ \citenamefont
  {Chen}(2021)}]{vonGersdorff21_metric_curvature}%
  \BibitemOpen
  \bibfield  {author} {\bibinfo {author} {\bibfnamefont {G.}~\bibnamefont {von
  Gersdorff}}\ and\ \bibinfo {author} {\bibfnamefont {W.}~\bibnamefont
  {Chen}},\ }\bibfield  {title} {\enquote {\bibinfo {title} {Measurement of
  topological order based on metric-curvature correspondence},}\ }\href
  {\doibase 10.1103/PhysRevB.104.195133} {\bibfield  {journal} {\bibinfo
  {journal} {Phys. Rev. B}\ }\textbf {\bibinfo {volume} {104}},\ \bibinfo
  {pages} {195133} (\bibinfo {year} {2021})}\BibitemShut {NoStop}%
\bibitem [{\citenamefont {Ghosh}\ \emph {et~al.}(2024)\citenamefont {Ghosh},
  \citenamefont {Onishi}, \citenamefont {Xu}, \citenamefont {Lin},
  \citenamefont {Fu},\ and\ \citenamefont {Bansil}}]{Ghosh24}%
  \BibitemOpen
  \bibfield  {author} {\bibinfo {author} {\bibfnamefont {B.}~\bibnamefont
  {Ghosh}}, \bibinfo {author} {\bibfnamefont {Y.}~\bibnamefont {Onishi}},
  \bibinfo {author} {\bibfnamefont {S.-Y.}\ \bibnamefont {Xu}}, \bibinfo
  {author} {\bibfnamefont {H.}~\bibnamefont {Lin}}, \bibinfo {author}
  {\bibfnamefont {L.}~\bibnamefont {Fu}}, \ and\ \bibinfo {author}
  {\bibfnamefont {A.}~\bibnamefont {Bansil}},\ }\bibfield  {title} {\enquote
  {\bibinfo {title} {Probing quantum geometry through optical conductivity and
  magnetic circular dichroism},}\ }\href {https://arxiv.org/abs/2401.09689} {\
  (\bibinfo {year} {2024})},\ \Eprint {http://arxiv.org/abs/2401.09689}
  {arXiv:2401.09689 [cond-mat.mtrl-sci]} \BibitemShut {NoStop}%
\bibitem [{\citenamefont {Ezawa}(2024)}]{Ezawa24}%
  \BibitemOpen
  \bibfield  {author} {\bibinfo {author} {\bibfnamefont {M.}~\bibnamefont
  {Ezawa}},\ }\bibfield  {title} {\enquote {\bibinfo {title} {Analytic approach
  to quantum metric and optical conductivity in dirac models with parabolic
  mass in arbitrary dimensions},}\ }\href {https://arxiv.org/abs/2408.02951} {\
   (\bibinfo {year} {2024})},\ \Eprint {http://arxiv.org/abs/2408.02951}
  {arXiv:2408.02951 [cond-mat.mes-hall]} \BibitemShut {NoStop}%
\bibitem [{\citenamefont {Komissarov}\ \emph {et~al.}(2024)\citenamefont
  {Komissarov}, \citenamefont {Holder},\ and\ \citenamefont
  {Queiroz}}]{Komissarov24}%
  \BibitemOpen
  \bibfield  {author} {\bibinfo {author} {\bibfnamefont {I.}~\bibnamefont
  {Komissarov}}, \bibinfo {author} {\bibfnamefont {T.}~\bibnamefont {Holder}},
  \ and\ \bibinfo {author} {\bibfnamefont {R.}~\bibnamefont {Queiroz}},\
  }\bibfield  {title} {\enquote {\bibinfo {title} {The quantum geometric origin
  of capacitance in insulators},}\ }\href {\doibase 10.1038/s41467-024-48808-x}
  {\bibfield  {journal} {\bibinfo  {journal} {Nature Communications}\ }\textbf
  {\bibinfo {volume} {15}},\ \bibinfo {pages} {4621} (\bibinfo {year}
  {2024})}\BibitemShut {NoStop}%
\bibitem [{\citenamefont {de~Sousa}\ \emph {et~al.}(2023)\citenamefont
  {de~Sousa}, \citenamefont {Cruz},\ and\ \citenamefont
  {Chen}}]{deSousa23_fidelity_marker}%
  \BibitemOpen
  \bibfield  {author} {\bibinfo {author} {\bibfnamefont {M.~S.~M.}\
  \bibnamefont {de~Sousa}}, \bibinfo {author} {\bibfnamefont {A.~L.}\
  \bibnamefont {Cruz}}, \ and\ \bibinfo {author} {\bibfnamefont
  {W.}~\bibnamefont {Chen}},\ }\bibfield  {title} {\enquote {\bibinfo {title}
  {Mapping quantum geometry and quantum phase transitions to real space by a
  fidelity marker},}\ }\href {\doibase 10.1103/PhysRevB.107.205133} {\bibfield
  {journal} {\bibinfo  {journal} {Phys. Rev. B}\ }\textbf {\bibinfo {volume}
  {107}},\ \bibinfo {pages} {205133} (\bibinfo {year} {2023})}\BibitemShut
  {NoStop}%
\bibitem [{\citenamefont {C\'ardenas-Castillo}\ \emph
  {et~al.}(2024)\citenamefont {C\'ardenas-Castillo}, \citenamefont {Zhang},
  \citenamefont {Freire}, \citenamefont {Kochan},\ and\ \citenamefont
  {Chen}}]{CardenasCastillo24_spread_Wannier}%
  \BibitemOpen
  \bibfield  {author} {\bibinfo {author} {\bibfnamefont {L.~F.}\ \bibnamefont
  {C\'ardenas-Castillo}}, \bibinfo {author} {\bibfnamefont {S.}~\bibnamefont
  {Zhang}}, \bibinfo {author} {\bibfnamefont {F.~L.}\ \bibnamefont {Freire}},
  \bibinfo {author} {\bibfnamefont {D.}~\bibnamefont {Kochan}}, \ and\ \bibinfo
  {author} {\bibfnamefont {W.}~\bibnamefont {Chen}},\ }\bibfield  {title}
  {\enquote {\bibinfo {title} {Detecting the spread of valence-band wannier
  functions by optical sum rules},}\ }\href {\doibase
  10.1103/PhysRevB.110.075203} {\bibfield  {journal} {\bibinfo  {journal}
  {Phys. Rev. B}\ }\textbf {\bibinfo {volume} {110}},\ \bibinfo {pages}
  {075203} (\bibinfo {year} {2024})}\BibitemShut {NoStop}%
\bibitem [{\citenamefont {Fox}(2010)}]{Fox10}%
  \BibitemOpen
  \bibfield  {author} {\bibinfo {author} {\bibfnamefont {M.}~\bibnamefont
  {Fox}},\ }\href@noop {} {\emph {\bibinfo {title} {Optical Properties of
  Solids}}}\ (\bibinfo  {publisher} {Oxford University Press, Oxford},\
  \bibinfo {year} {2010})\BibitemShut {NoStop}%
\bibitem [{\citenamefont {Tanner}(2019)}]{Tanner19}%
  \BibitemOpen
  \bibfield  {author} {\bibinfo {author} {\bibfnamefont {D.~B.}\ \bibnamefont
  {Tanner}},\ }\href@noop {} {\emph {\bibinfo {title} {Optical Effects in
  Solids}}}\ (\bibinfo  {publisher} {Cambridge University Press, Cambridge},\
  \bibinfo {year} {2019})\BibitemShut {NoStop}%
\bibitem [{\citenamefont {Bianco}\ and\ \citenamefont
  {Resta}(2011)}]{Bianco11}%
  \BibitemOpen
  \bibfield  {author} {\bibinfo {author} {\bibfnamefont {R.}~\bibnamefont
  {Bianco}}\ and\ \bibinfo {author} {\bibfnamefont {R.}~\bibnamefont {Resta}},\
  }\bibfield  {title} {\enquote {\bibinfo {title} {Mapping topological order in
  coordinate space},}\ }\href {\doibase 10.1103/PhysRevB.84.241106} {\bibfield
  {journal} {\bibinfo  {journal} {Phys. Rev. B}\ }\textbf {\bibinfo {volume}
  {84}},\ \bibinfo {pages} {241106} (\bibinfo {year} {2011})}\BibitemShut
  {NoStop}%
\bibitem [{\citenamefont {Prodan}\ \emph {et~al.}(2010)\citenamefont {Prodan},
  \citenamefont {Hughes},\ and\ \citenamefont {Bernevig}}]{Prodan10}%
  \BibitemOpen
  \bibfield  {author} {\bibinfo {author} {\bibfnamefont {E.}~\bibnamefont
  {Prodan}}, \bibinfo {author} {\bibfnamefont {T.~L.}\ \bibnamefont {Hughes}},
  \ and\ \bibinfo {author} {\bibfnamefont {B.~A.}\ \bibnamefont {Bernevig}},\
  }\bibfield  {title} {\enquote {\bibinfo {title} {Entanglement spectrum of a
  disordered topological chern insulator},}\ }\href {\doibase
  10.1103/PhysRevLett.105.115501} {\bibfield  {journal} {\bibinfo  {journal}
  {Phys. Rev. Lett.}\ }\textbf {\bibinfo {volume} {105}},\ \bibinfo {pages}
  {115501} (\bibinfo {year} {2010})}\BibitemShut {NoStop}%
\bibitem [{\citenamefont {Prodan}(2010)}]{Prodan10_2}%
  \BibitemOpen
  \bibfield  {author} {\bibinfo {author} {\bibfnamefont {E.}~\bibnamefont
  {Prodan}},\ }\bibfield  {title} {\enquote {\bibinfo {title} {Non-commutative
  tools for topological insulators},}\ }\href {\doibase
  10.1088/1367-2630/12/6/065003} {\bibfield  {journal} {\bibinfo  {journal}
  {New J. Phys.}\ }\textbf {\bibinfo {volume} {12}},\ \bibinfo {pages} {065003}
  (\bibinfo {year} {2010})}\BibitemShut {NoStop}%
\bibitem [{\citenamefont {Prodan}(2011)}]{Prodan11}%
  \BibitemOpen
  \bibfield  {author} {\bibinfo {author} {\bibfnamefont {E.}~\bibnamefont
  {Prodan}},\ }\bibfield  {title} {\enquote {\bibinfo {title} {Disordered
  topological insulators: a non-commutative geometry perspective},}\ }\href
  {\doibase 10.1088/1751-8113/44/11/113001} {\bibfield  {journal} {\bibinfo
  {journal} {J. Phys. A: Math. Theor.}\ }\textbf {\bibinfo {volume} {44}},\
  \bibinfo {pages} {113001} (\bibinfo {year} {2011})}\BibitemShut {NoStop}%
\bibitem [{\citenamefont {Chen}(2023)}]{Chen23_universal_marker}%
  \BibitemOpen
  \bibfield  {author} {\bibinfo {author} {\bibfnamefont {W.}~\bibnamefont
  {Chen}},\ }\bibfield  {title} {\enquote {\bibinfo {title} {Universal
  topological marker},}\ }\href {\doibase 10.1103/PhysRevB.107.045111}
  {\bibfield  {journal} {\bibinfo  {journal} {Phys. Rev. B}\ }\textbf {\bibinfo
  {volume} {107}},\ \bibinfo {pages} {045111} (\bibinfo {year}
  {2023})}\BibitemShut {NoStop}%
\bibitem [{\citenamefont {Marrazzo}\ and\ \citenamefont
  {Resta}(2019)}]{Marrazzo19}%
  \BibitemOpen
  \bibfield  {author} {\bibinfo {author} {\bibfnamefont {A.}~\bibnamefont
  {Marrazzo}}\ and\ \bibinfo {author} {\bibfnamefont {R.}~\bibnamefont
  {Resta}},\ }\bibfield  {title} {\enquote {\bibinfo {title} {Local theory of
  the insulating state},}\ }\href {\doibase 10.1103/PhysRevLett.122.166602}
  {\bibfield  {journal} {\bibinfo  {journal} {Phys. Rev. Lett.}\ }\textbf
  {\bibinfo {volume} {122}},\ \bibinfo {pages} {166602} (\bibinfo {year}
  {2019})}\BibitemShut {NoStop}%
\bibitem [{\citenamefont {Zhang}\ \emph {et~al.}(2009)\citenamefont {Zhang},
  \citenamefont {Liu}, \citenamefont {Qi}, \citenamefont {Dai}, \citenamefont
  {Fang},\ and\ \citenamefont {Zhang}}]{Zhang09}%
  \BibitemOpen
  \bibfield  {author} {\bibinfo {author} {\bibfnamefont {H.}~\bibnamefont
  {Zhang}}, \bibinfo {author} {\bibfnamefont {C.~X.}\ \bibnamefont {Liu}},
  \bibinfo {author} {\bibfnamefont {X.~L.}\ \bibnamefont {Qi}}, \bibinfo
  {author} {\bibfnamefont {X.}~\bibnamefont {Dai}}, \bibinfo {author}
  {\bibfnamefont {Z.}~\bibnamefont {Fang}}, \ and\ \bibinfo {author}
  {\bibfnamefont {S.~C.}\ \bibnamefont {Zhang}},\ }\bibfield  {title} {\enquote
  {\bibinfo {title} {Topological insulators in bi2se3, bi2te3 and sb2te3 with a
  single dirac cone on the surface},}\ }\href {\doibase 10.1038/nphys1270}
  {\bibfield  {journal} {\bibinfo  {journal} {Nat. Phys.}\ }\textbf {\bibinfo
  {volume} {5}},\ \bibinfo {pages} {438} (\bibinfo {year} {2009})}\BibitemShut
  {NoStop}%
\bibitem [{\citenamefont {Liu}\ \emph {et~al.}(2010)\citenamefont {Liu},
  \citenamefont {Qi}, \citenamefont {Zhang}, \citenamefont {Dai}, \citenamefont
  {Fang},\ and\ \citenamefont {Zhang}}]{Liu10}%
  \BibitemOpen
  \bibfield  {author} {\bibinfo {author} {\bibfnamefont {C.~X.}\ \bibnamefont
  {Liu}}, \bibinfo {author} {\bibfnamefont {X.~L.}\ \bibnamefont {Qi}},
  \bibinfo {author} {\bibfnamefont {H.}~\bibnamefont {Zhang}}, \bibinfo
  {author} {\bibfnamefont {X.}~\bibnamefont {Dai}}, \bibinfo {author}
  {\bibfnamefont {Z.}~\bibnamefont {Fang}}, \ and\ \bibinfo {author}
  {\bibfnamefont {S.~C.}\ \bibnamefont {Zhang}},\ }\bibfield  {title} {\enquote
  {\bibinfo {title} {Model hamiltonian for topological insulators},}\ }\href
  {\doibase 10.1103/PhysRevB.82.045122} {\bibfield  {journal} {\bibinfo
  {journal} {Phys. Rev. B}\ }\textbf {\bibinfo {volume} {82}},\ \bibinfo
  {pages} {045122} (\bibinfo {year} {2010})}\BibitemShut {NoStop}%
\bibitem [{\citenamefont {Matsuura}\ and\ \citenamefont
  {Ryu}(2010)}]{Matsuura10}%
  \BibitemOpen
  \bibfield  {author} {\bibinfo {author} {\bibfnamefont {S.}~\bibnamefont
  {Matsuura}}\ and\ \bibinfo {author} {\bibfnamefont {S.}~\bibnamefont {Ryu}},\
  }\bibfield  {title} {\enquote {\bibinfo {title} {Momentum space metric,
  nonlocal operator, and topological insulators},}\ }\href {\doibase
  10.1103/PhysRevB.82.245113} {\bibfield  {journal} {\bibinfo  {journal} {Phys.
  Rev. B}\ }\textbf {\bibinfo {volume} {82}},\ \bibinfo {pages} {245113}
  (\bibinfo {year} {2010})}\BibitemShut {NoStop}%
\bibitem [{\citenamefont {Mera}\ \emph {et~al.}(2022)\citenamefont {Mera},
  \citenamefont {Zhang},\ and\ \citenamefont {Goldman}}]{Mera22}%
  \BibitemOpen
  \bibfield  {author} {\bibinfo {author} {\bibfnamefont {B.}~\bibnamefont
  {Mera}}, \bibinfo {author} {\bibfnamefont {A.}~\bibnamefont {Zhang}}, \ and\
  \bibinfo {author} {\bibfnamefont {N.}~\bibnamefont {Goldman}},\ }\bibfield
  {title} {\enquote {\bibinfo {title} {{Relating the topology of Dirac
  Hamiltonians to quantum geometry: When the quantum metric dictates Chern
  numbers and winding numbers}},}\ }\href {\doibase
  10.21468/SciPostPhys.12.1.018} {\bibfield  {journal} {\bibinfo  {journal}
  {SciPost Phys.}\ }\textbf {\bibinfo {volume} {12}},\ \bibinfo {pages} {018}
  (\bibinfo {year} {2022})}\BibitemShut {NoStop}%
\bibitem [{\citenamefont {Schnyder}\ \emph {et~al.}(2008)\citenamefont
  {Schnyder}, \citenamefont {Ryu}, \citenamefont {Furusaki},\ and\
  \citenamefont {Ludwig}}]{Schnyder08}%
  \BibitemOpen
  \bibfield  {author} {\bibinfo {author} {\bibfnamefont {A.~P.}\ \bibnamefont
  {Schnyder}}, \bibinfo {author} {\bibfnamefont {S.}~\bibnamefont {Ryu}},
  \bibinfo {author} {\bibfnamefont {A.}~\bibnamefont {Furusaki}}, \ and\
  \bibinfo {author} {\bibfnamefont {A.~W.~W.}\ \bibnamefont {Ludwig}},\
  }\bibfield  {title} {\enquote {\bibinfo {title} {Classification of
  topological insulators and superconductors in three spatial dimensions},}\
  }\href {\doibase 10.1103/PhysRevB.78.195125} {\bibfield  {journal} {\bibinfo
  {journal} {Phys. Rev. B}\ }\textbf {\bibinfo {volume} {78}},\ \bibinfo
  {pages} {195125} (\bibinfo {year} {2008})}\BibitemShut {NoStop}%
\bibitem [{\citenamefont {Ryu}\ \emph {et~al.}(2010)\citenamefont {Ryu},
  \citenamefont {Schnyder}, \citenamefont {Furusaki},\ and\ \citenamefont
  {Ludwig}}]{Ryu10}%
  \BibitemOpen
  \bibfield  {author} {\bibinfo {author} {\bibfnamefont {S.}~\bibnamefont
  {Ryu}}, \bibinfo {author} {\bibfnamefont {A.~P.}\ \bibnamefont {Schnyder}},
  \bibinfo {author} {\bibfnamefont {A.}~\bibnamefont {Furusaki}}, \ and\
  \bibinfo {author} {\bibfnamefont {A.~W.~W.}\ \bibnamefont {Ludwig}},\
  }\bibfield  {title} {\enquote {\bibinfo {title} {Topological insulators and
  superconductors: tenfold way and dimensional hierarchy},}\ }\href
  {http://stacks.iop.org/1367-2630/12/i=6/a=065010} {\bibfield  {journal}
  {\bibinfo  {journal} {New J. Phys.}\ }\textbf {\bibinfo {volume} {12}},\
  \bibinfo {pages} {065010} (\bibinfo {year} {2010})}\BibitemShut {NoStop}%
\bibitem [{\citenamefont {Chiu}\ \emph {et~al.}(2016)\citenamefont {Chiu},
  \citenamefont {Teo}, \citenamefont {Schnyder},\ and\ \citenamefont
  {Ryu}}]{Chiu16}%
  \BibitemOpen
  \bibfield  {author} {\bibinfo {author} {\bibfnamefont {C.~K.}\ \bibnamefont
  {Chiu}}, \bibinfo {author} {\bibfnamefont {J.~C.~Y.}\ \bibnamefont {Teo}},
  \bibinfo {author} {\bibfnamefont {A.~P.}\ \bibnamefont {Schnyder}}, \ and\
  \bibinfo {author} {\bibfnamefont {S.}~\bibnamefont {Ryu}},\ }\bibfield
  {title} {\enquote {\bibinfo {title} {Classification of topological quantum
  matter with symmetries},}\ }\href {\doibase 10.1103/RevModPhys.88.035005}
  {\bibfield  {journal} {\bibinfo  {journal} {Rev. Mod. Phys.}\ }\textbf
  {\bibinfo {volume} {88}},\ \bibinfo {pages} {035005} (\bibinfo {year}
  {2016})}\BibitemShut {NoStop}%
\bibitem [{\citenamefont {Boris I.~Shklovskii}(1984)}]{Shklovskii84}%
  \BibitemOpen
  \bibfield  {author} {\bibinfo {author} {\bibfnamefont {A.~L.~E.}\
  \bibnamefont {Boris I.~Shklovskii}},\ }\href@noop {} {\emph {\bibinfo {title}
  {Electronic Properties of Doped Semiconductors}}}\ (\bibinfo  {publisher}
  {Springer-Verlag, Berlin},\ \bibinfo {year} {1984})\BibitemShut {NoStop}%
\bibitem [{\citenamefont {Kirkpatrick}(1973)}]{Kirkpatrick73}%
  \BibitemOpen
  \bibfield  {author} {\bibinfo {author} {\bibfnamefont {S.}~\bibnamefont
  {Kirkpatrick}},\ }\bibfield  {title} {\enquote {\bibinfo {title} {Percolation
  and conduction},}\ }\href {\doibase 10.1103/RevModPhys.45.574} {\bibfield
  {journal} {\bibinfo  {journal} {Rev. Mod. Phys.}\ }\textbf {\bibinfo {volume}
  {45}},\ \bibinfo {pages} {574--588} (\bibinfo {year} {1973})}\BibitemShut
  {NoStop}%
\bibitem [{\citenamefont {Stroud}(1998)}]{Stroud98}%
  \BibitemOpen
  \bibfield  {author} {\bibinfo {author} {\bibfnamefont {D.}~\bibnamefont
  {Stroud}},\ }\bibfield  {title} {\enquote {\bibinfo {title} {The effective
  medium approximations: Some recent developments},}\ }\href {\doibase
  https://doi.org/10.1006/spmi.1997.0524} {\bibfield  {journal} {\bibinfo
  {journal} {Superlattices Microstruct.}\ }\textbf {\bibinfo {volume} {23}},\
  \bibinfo {pages} {567--573} (\bibinfo {year} {1998})}\BibitemShut {NoStop}%
\bibitem [{\citenamefont {Sihvola}(1999)}]{Sihvola99}%
  \BibitemOpen
  \bibfield  {author} {\bibinfo {author} {\bibfnamefont {A.}~\bibnamefont
  {Sihvola}},\ }\href@noop {} {\emph {\bibinfo {title} {Electromagnetic Mixing
  Formulas and Applications}}}\ (\bibinfo  {publisher} {Institution of
  Engineering and Technology, London},\ \bibinfo {year} {1999})\BibitemShut
  {NoStop}%
\bibitem [{\citenamefont {Yu}\ and\ \citenamefont {Cardona}(2010)}]{Yu10}%
  \BibitemOpen
  \bibfield  {author} {\bibinfo {author} {\bibfnamefont {P.}~\bibnamefont
  {Yu}}\ and\ \bibinfo {author} {\bibfnamefont {M.}~\bibnamefont {Cardona}},\
  }\href@noop {} {\emph {\bibinfo {title} {Fundamentals of Semiconductors:
  Physics and Materials Properties}}}\ (\bibinfo  {publisher} {Springer,
  Berlin},\ \bibinfo {year} {2010})\BibitemShut {NoStop}%
\bibitem [{\citenamefont {Marzari}\ and\ \citenamefont
  {Vanderbilt}(1997)}]{Marzari97}%
  \BibitemOpen
  \bibfield  {author} {\bibinfo {author} {\bibfnamefont {N.}~\bibnamefont
  {Marzari}}\ and\ \bibinfo {author} {\bibfnamefont {D.}~\bibnamefont
  {Vanderbilt}},\ }\bibfield  {title} {\enquote {\bibinfo {title} {Maximally
  localized generalized wannier functions for composite energy bands},}\ }\href
  {\doibase 10.1103/PhysRevB.56.12847} {\bibfield  {journal} {\bibinfo
  {journal} {Phys. Rev. B}\ }\textbf {\bibinfo {volume} {56}},\ \bibinfo
  {pages} {12847--12865} (\bibinfo {year} {1997})}\BibitemShut {NoStop}%
\bibitem [{\citenamefont {Marzari}\ \emph {et~al.}(2012)\citenamefont
  {Marzari}, \citenamefont {Mostofi}, \citenamefont {Yates}, \citenamefont
  {Souza},\ and\ \citenamefont {Vanderbilt}}]{Marzari12}%
  \BibitemOpen
  \bibfield  {author} {\bibinfo {author} {\bibfnamefont {N.}~\bibnamefont
  {Marzari}}, \bibinfo {author} {\bibfnamefont {A.~A.}\ \bibnamefont
  {Mostofi}}, \bibinfo {author} {\bibfnamefont {J.~R.}\ \bibnamefont {Yates}},
  \bibinfo {author} {\bibfnamefont {I.}~\bibnamefont {Souza}}, \ and\ \bibinfo
  {author} {\bibfnamefont {D.}~\bibnamefont {Vanderbilt}},\ }\bibfield  {title}
  {\enquote {\bibinfo {title} {Maximally localized wannier functions: Theory
  and applications},}\ }\href {\doibase 10.1103/RevModPhys.84.1419} {\bibfield
  {journal} {\bibinfo  {journal} {Rev. Mod. Phys.}\ }\textbf {\bibinfo {volume}
  {84}},\ \bibinfo {pages} {1419--1475} (\bibinfo {year} {2012})}\BibitemShut
  {NoStop}%
\bibitem [{\citenamefont {Molignini}\ \emph {et~al.}(2023)\citenamefont
  {Molignini}, \citenamefont {Lapierre}, \citenamefont {Chitra},\ and\
  \citenamefont {Chen}}]{Molignini23_Chern_marker}%
  \BibitemOpen
  \bibfield  {author} {\bibinfo {author} {\bibfnamefont {P.}~\bibnamefont
  {Molignini}}, \bibinfo {author} {\bibfnamefont {B.}~\bibnamefont {Lapierre}},
  \bibinfo {author} {\bibfnamefont {R.}~\bibnamefont {Chitra}}, \ and\ \bibinfo
  {author} {\bibfnamefont {W.}~\bibnamefont {Chen}},\ }\bibfield  {title}
  {\enquote {\bibinfo {title} {{Probing Chern number by opacity and topological
  phase transition by a nonlocal Chern marker}},}\ }\href {\doibase
  10.21468/SciPostPhysCore.6.3.059} {\bibfield  {journal} {\bibinfo  {journal}
  {SciPost Phys. Core}\ }\textbf {\bibinfo {volume} {6}},\ \bibinfo {pages}
  {059} (\bibinfo {year} {2023})}\BibitemShut {NoStop}%
\bibitem [{\citenamefont {Betzig}\ \emph {et~al.}(1986)\citenamefont {Betzig},
  \citenamefont {Lewis}, \citenamefont {Harootunian}, \citenamefont
  {Isaacson},\ and\ \citenamefont {Kratschmer}}]{Betzig86}%
  \BibitemOpen
  \bibfield  {author} {\bibinfo {author} {\bibfnamefont {E.}~\bibnamefont
  {Betzig}}, \bibinfo {author} {\bibfnamefont {A.}~\bibnamefont {Lewis}},
  \bibinfo {author} {\bibfnamefont {A.}~\bibnamefont {Harootunian}}, \bibinfo
  {author} {\bibfnamefont {M.}~\bibnamefont {Isaacson}}, \ and\ \bibinfo
  {author} {\bibfnamefont {E.}~\bibnamefont {Kratschmer}},\ }\bibfield  {title}
  {\enquote {\bibinfo {title} {Near field scanning optical microscopy (nsom):
  Development and biophysical applications},}\ }\href {\doibase
  10.1016/S0006-3495(86)83640-2} {\bibfield  {journal} {\bibinfo  {journal}
  {Biophys. J.}\ }\textbf {\bibinfo {volume} {49}},\ \bibinfo {pages}
  {269--279} (\bibinfo {year} {1986})}\BibitemShut {NoStop}%
\bibitem [{\citenamefont {Hsu}(2001)}]{Hsu01}%
  \BibitemOpen
  \bibfield  {author} {\bibinfo {author} {\bibfnamefont {J.~W.~P.}\
  \bibnamefont {Hsu}},\ }\bibfield  {title} {\enquote {\bibinfo {title}
  {Near-field scanning optical microscopy studies of electronic and photonic
  materials and devices},}\ }\href {\doibase
  https://doi.org/10.1016/S0927-796X(00)00031-0} {\bibfield  {journal}
  {\bibinfo  {journal} {Mater. Sci. Eng. R: Rep.}\ }\textbf {\bibinfo {volume}
  {33}},\ \bibinfo {pages} {1--50} (\bibinfo {year} {2001})}\BibitemShut
  {NoStop}%
\bibitem [{\citenamefont {Kim}\ and\ \citenamefont {Song}(2007)}]{Kim07}%
  \BibitemOpen
  \bibfield  {author} {\bibinfo {author} {\bibfnamefont {J.~H.}\ \bibnamefont
  {Kim}}\ and\ \bibinfo {author} {\bibfnamefont {K.-B.}\ \bibnamefont {Song}},\
  }\bibfield  {title} {\enquote {\bibinfo {title} {Recent progress of
  nano-technology with nsom},}\ }\href {\doibase
  https://doi.org/10.1016/j.micron.2006.06.010} {\bibfield  {journal} {\bibinfo
   {journal} {Micron}\ }\textbf {\bibinfo {volume} {38}},\ \bibinfo {pages}
  {409--426} (\bibinfo {year} {2007})}\BibitemShut {NoStop}%
\bibitem [{\citenamefont {Tranca}\ \emph {et~al.}(2018)\citenamefont {Tranca},
  \citenamefont {Stanciu}, \citenamefont {Hristu}, \citenamefont {Witgen},\
  and\ \citenamefont {Stanciu}}]{Tranca18}%
  \BibitemOpen
  \bibfield  {author} {\bibinfo {author} {\bibfnamefont {Denis~E.}\
  \bibnamefont {Tranca}}, \bibinfo {author} {\bibfnamefont {Stefan~G.}\
  \bibnamefont {Stanciu}}, \bibinfo {author} {\bibfnamefont {Radu}\
  \bibnamefont {Hristu}}, \bibinfo {author} {\bibfnamefont {Brent~M.}\
  \bibnamefont {Witgen}}, \ and\ \bibinfo {author} {\bibfnamefont {George~A.}\
  \bibnamefont {Stanciu}},\ }\bibfield  {title} {\enquote {\bibinfo {title}
  {Nanoscale mapping of refractive index by using scattering-type scanning
  near-field optical microscopy},}\ }\href {\doibase
  https://doi.org/10.1016/j.nano.2017.08.016} {\bibfield  {journal} {\bibinfo
  {journal} {NBM}\ }\textbf {\bibinfo {volume} {14}},\ \bibinfo {pages}
  {47--50} (\bibinfo {year} {2018})}\BibitemShut {NoStop}%
\bibitem [{\citenamefont {Souza}\ \emph {et~al.}(2000)\citenamefont {Souza},
  \citenamefont {Wilkens},\ and\ \citenamefont {Martin}}]{Souza00}%
  \BibitemOpen
  \bibfield  {author} {\bibinfo {author} {\bibfnamefont {I.}~\bibnamefont
  {Souza}}, \bibinfo {author} {\bibfnamefont {T.}~\bibnamefont {Wilkens}}, \
  and\ \bibinfo {author} {\bibfnamefont {R.~M.}\ \bibnamefont {Martin}},\
  }\bibfield  {title} {\enquote {\bibinfo {title} {Polarization and
  localization in insulators: Generating function approach},}\ }\href {\doibase
  10.1103/PhysRevB.62.1666} {\bibfield  {journal} {\bibinfo  {journal} {Phys.
  Rev. B}\ }\textbf {\bibinfo {volume} {62}},\ \bibinfo {pages} {1666--1683}
  (\bibinfo {year} {2000})}\BibitemShut {NoStop}%
\bibitem [{\citenamefont {Reed}\ \emph {et~al.}(1999)\citenamefont {Reed},
  \citenamefont {Chen}, \citenamefont {MacDonald}, \citenamefont {Silcox},\
  and\ \citenamefont {Bertsch}}]{Reed99}%
  \BibitemOpen
  \bibfield  {author} {\bibinfo {author} {\bibfnamefont {B.~W.}\ \bibnamefont
  {Reed}}, \bibinfo {author} {\bibfnamefont {J.~M.}\ \bibnamefont {Chen}},
  \bibinfo {author} {\bibfnamefont {N.~C.}\ \bibnamefont {MacDonald}}, \bibinfo
  {author} {\bibfnamefont {J.}~\bibnamefont {Silcox}}, \ and\ \bibinfo {author}
  {\bibfnamefont {G.~F.}\ \bibnamefont {Bertsch}},\ }\bibfield  {title}
  {\enquote {\bibinfo {title} {Fabrication and stem/eels measurements of
  nanometer-scale silicon tips and filaments},}\ }\href {\doibase
  10.1103/PhysRevB.60.5641} {\bibfield  {journal} {\bibinfo  {journal} {Phys.
  Rev. B}\ }\textbf {\bibinfo {volume} {60}},\ \bibinfo {pages} {5641--5652}
  (\bibinfo {year} {1999})}\BibitemShut {NoStop}%
\bibitem [{\citenamefont {Madelung}(2004)}]{Madelung04}%
  \BibitemOpen
  \bibfield  {author} {\bibinfo {author} {\bibfnamefont {O.}~\bibnamefont
  {Madelung}},\ }\href@noop {} {\emph {\bibinfo {title} {Semiconductors: Data
  Handbook}}}\ (\bibinfo  {publisher} {Springer, Heidelberg},\ \bibinfo {year}
  {2004})\BibitemShut {NoStop}%
\bibitem [{\citenamefont {Greenaway}\ and\ \citenamefont
  {Harbeke}(1965)}]{Greenaway65}%
  \BibitemOpen
  \bibfield  {author} {\bibinfo {author} {\bibfnamefont {D.L.}\ \bibnamefont
  {Greenaway}}\ and\ \bibinfo {author} {\bibfnamefont {G.}~\bibnamefont
  {Harbeke}},\ }\bibfield  {title} {\enquote {\bibinfo {title} {Band structure
  of bismuth telluride, bismuth selenide and their respective alloys},}\ }\href
  {\doibase https://doi.org/10.1016/0022-3697(65)90092-2} {\bibfield  {journal}
  {\bibinfo  {journal} {J. Phys. Chem. Solids}\ }\textbf {\bibinfo {volume}
  {26}},\ \bibinfo {pages} {1585--1604} (\bibinfo {year} {1965})}\BibitemShut
  {NoStop}%
\bibitem [{\citenamefont {Oliveira}\ and\ \citenamefont
  {Chen}(2024)}]{Oliveira24_impurity_marker}%
  \BibitemOpen
  \bibfield  {author} {\bibinfo {author} {\bibfnamefont {L.~A.}\ \bibnamefont
  {Oliveira}}\ and\ \bibinfo {author} {\bibfnamefont {W.}~\bibnamefont
  {Chen}},\ }\bibfield  {title} {\enquote {\bibinfo {title} {Robustness of
  topological order against disorder},}\ }\href {\doibase
  10.1103/PhysRevB.109.094202} {\bibfield  {journal} {\bibinfo  {journal}
  {Phys. Rev. B}\ }\textbf {\bibinfo {volume} {109}},\ \bibinfo {pages}
  {094202} (\bibinfo {year} {2024})}\BibitemShut {NoStop}%
\bibitem [{\citenamefont {Chen}(2020)}]{Chen20_absence_edge_current}%
  \BibitemOpen
  \bibfield  {author} {\bibinfo {author} {\bibfnamefont {W.}~\bibnamefont
  {Chen}},\ }\bibfield  {title} {\enquote {\bibinfo {title} {Absence of
  equilibrium edge currents in theoretical models of topological insulators},}\
  }\href {\doibase 10.1103/PhysRevB.101.195120} {\bibfield  {journal} {\bibinfo
   {journal} {Phys. Rev. B}\ }\textbf {\bibinfo {volume} {101}},\ \bibinfo
  {pages} {195120} (\bibinfo {year} {2020})}\BibitemShut {NoStop}%
\bibitem [{\citenamefont {Ou}\ \emph {et~al.}(2014)\citenamefont {Ou},
  \citenamefont {So}, \citenamefont {Adamo}, \citenamefont {Sulaev},
  \citenamefont {Wang},\ and\ \citenamefont {Zheludev}}]{Ou14}%
  \BibitemOpen
  \bibfield  {author} {\bibinfo {author} {\bibfnamefont {J.~Y.}\ \bibnamefont
  {Ou}}, \bibinfo {author} {\bibfnamefont {J.~K.}\ \bibnamefont {So}}, \bibinfo
  {author} {\bibfnamefont {G.}~\bibnamefont {Adamo}}, \bibinfo {author}
  {\bibfnamefont {A.}~\bibnamefont {Sulaev}}, \bibinfo {author} {\bibfnamefont
  {L.}~\bibnamefont {Wang}}, \ and\ \bibinfo {author} {\bibfnamefont {N.~I.}\
  \bibnamefont {Zheludev}},\ }\bibfield  {title} {\enquote {\bibinfo {title}
  {Ultraviolet and visible range plasmonics in the topological insulator
  bi1.5sb0.5te1.8se1.2},}\ }\href {\doibase 10.1038/ncomms6139} {\bibfield
  {journal} {\bibinfo  {journal} {Nat. Commun.}\ }\textbf {\bibinfo {volume}
  {5}},\ \bibinfo {pages} {5139} (\bibinfo {year} {2014})}\BibitemShut
  {NoStop}%
\bibitem [{\citenamefont {Zhao}\ \emph {et~al.}(2015)\citenamefont {Zhao},
  \citenamefont {Bosman}, \citenamefont {Danesh}, \citenamefont {Zeng},
  \citenamefont {Song}, \citenamefont {Darma}, \citenamefont {Rusydi},
  \citenamefont {Lin}, \citenamefont {Qiu},\ and\ \citenamefont
  {Loh}}]{Zhao15}%
  \BibitemOpen
  \bibfield  {author} {\bibinfo {author} {\bibfnamefont {M.}~\bibnamefont
  {Zhao}}, \bibinfo {author} {\bibfnamefont {M.}~\bibnamefont {Bosman}},
  \bibinfo {author} {\bibfnamefont {M.}~\bibnamefont {Danesh}}, \bibinfo
  {author} {\bibfnamefont {M.}~\bibnamefont {Zeng}}, \bibinfo {author}
  {\bibfnamefont {P.}~\bibnamefont {Song}}, \bibinfo {author} {\bibfnamefont
  {Y.}~\bibnamefont {Darma}}, \bibinfo {author} {\bibfnamefont
  {A.}~\bibnamefont {Rusydi}}, \bibinfo {author} {\bibfnamefont
  {H.}~\bibnamefont {Lin}}, \bibinfo {author} {\bibfnamefont {C.~W.}\
  \bibnamefont {Qiu}}, \ and\ \bibinfo {author} {\bibfnamefont {K.~P.}\
  \bibnamefont {Loh}},\ }\bibfield  {title} {\enquote {\bibinfo {title}
  {Visible surface plasmon modes in single bi2te3 nanoplate},}\ }\href
  {\doibase 10.1021/acs.nanolett.5b03966} {\bibfield  {journal} {\bibinfo
  {journal} {Nano Lett.}\ }\textbf {\bibinfo {volume} {15}},\ \bibinfo {pages}
  {8331--8335} (\bibinfo {year} {2015})}\BibitemShut {NoStop}%
\bibitem [{\citenamefont {Yin}\ \emph {et~al.}(2017)\citenamefont {Yin},
  \citenamefont {Krishnamoorthy}, \citenamefont {Adamo}, \citenamefont
  {Dubrovkin}, \citenamefont {Chong}, \citenamefont {Zheludev},\ and\
  \citenamefont {Soci}}]{Yin17}%
  \BibitemOpen
  \bibfield  {author} {\bibinfo {author} {\bibfnamefont {J.}~\bibnamefont
  {Yin}}, \bibinfo {author} {\bibfnamefont {H.~N.}\ \bibnamefont
  {Krishnamoorthy}}, \bibinfo {author} {\bibfnamefont {G.}~\bibnamefont
  {Adamo}}, \bibinfo {author} {\bibfnamefont {A.~M.}\ \bibnamefont
  {Dubrovkin}}, \bibinfo {author} {\bibfnamefont {Y.}~\bibnamefont {Chong}},
  \bibinfo {author} {\bibfnamefont {N.~I.}\ \bibnamefont {Zheludev}}, \ and\
  \bibinfo {author} {\bibfnamefont {C.}~\bibnamefont {Soci}},\ }\bibfield
  {title} {\enquote {\bibinfo {title} {Plasmonics of topological insulators at
  optical frequencies},}\ }\href {\doibase 10.1038/am.2017.149} {\bibfield
  {journal} {\bibinfo  {journal} {NPG Asia Mater.}\ }\textbf {\bibinfo {volume}
  {9}},\ \bibinfo {pages} {e425--e425} (\bibinfo {year} {2017})}\BibitemShut
  {NoStop}%
\bibitem [{\citenamefont {Chen}\ and\ \citenamefont {von
  Gersdorff}(2022)}]{Chen22_dressed_Berry_metric}%
  \BibitemOpen
  \bibfield  {author} {\bibinfo {author} {\bibfnamefont {Wei}\ \bibnamefont
  {Chen}}\ and\ \bibinfo {author} {\bibfnamefont {Gero}\ \bibnamefont {von
  Gersdorff}},\ }\bibfield  {title} {\enquote {\bibinfo {title} {Measurement of
  interaction-dressed berry curvature and quantum metric in solids by optical
  absorption},}\ }\href {\doibase 10.21468/SciPostPhysCore.5.3.040} {\bibfield
  {journal} {\bibinfo  {journal} {SciPost Phys. Core}\ }\textbf {\bibinfo
  {volume} {5}},\ \bibinfo {pages} {040} (\bibinfo {year} {2022})}\BibitemShut
  {NoStop}%
\bibitem [{\citenamefont {Hosur}\ \emph {et~al.}(2012)\citenamefont {Hosur},
  \citenamefont {Parameswaran},\ and\ \citenamefont {Vishwanath}}]{Hosur12}%
  \BibitemOpen
  \bibfield  {author} {\bibinfo {author} {\bibfnamefont {P.}~\bibnamefont
  {Hosur}}, \bibinfo {author} {\bibfnamefont {S.~A.}\ \bibnamefont
  {Parameswaran}}, \ and\ \bibinfo {author} {\bibfnamefont {A.}~\bibnamefont
  {Vishwanath}},\ }\bibfield  {title} {\enquote {\bibinfo {title} {Charge
  transport in weyl semimetals},}\ }\href {\doibase
  10.1103/PhysRevLett.108.046602} {\bibfield  {journal} {\bibinfo  {journal}
  {Phys. Rev. Lett.}\ }\textbf {\bibinfo {volume} {108}},\ \bibinfo {pages}
  {046602} (\bibinfo {year} {2012})}\BibitemShut {NoStop}%
\bibitem [{\citenamefont {B\'acsi}\ and\ \citenamefont
  {Virosztek}(2013)}]{Bacsi13}%
  \BibitemOpen
  \bibfield  {author} {\bibinfo {author} {\bibfnamefont {\.}~\bibnamefont
  {B\'acsi}}\ and\ \bibinfo {author} {\bibfnamefont {A.}~\bibnamefont
  {Virosztek}},\ }\bibfield  {title} {\enquote {\bibinfo {title} {Low-frequency
  optical conductivity in graphene and in other scale-invariant two-band
  systems},}\ }\href {\doibase 10.1103/PhysRevB.87.125425} {\bibfield
  {journal} {\bibinfo  {journal} {Phys. Rev. B}\ }\textbf {\bibinfo {volume}
  {87}},\ \bibinfo {pages} {125425} (\bibinfo {year} {2013})}\BibitemShut
  {NoStop}%
\bibitem [{\citenamefont {Timusk}\ \emph {et~al.}(2013)\citenamefont {Timusk},
  \citenamefont {Carbotte}, \citenamefont {Homes}, \citenamefont {Basov},\ and\
  \citenamefont {Sharapov}}]{Timusk13}%
  \BibitemOpen
  \bibfield  {author} {\bibinfo {author} {\bibfnamefont {T.}~\bibnamefont
  {Timusk}}, \bibinfo {author} {\bibfnamefont {J.~P.}\ \bibnamefont
  {Carbotte}}, \bibinfo {author} {\bibfnamefont {C.~C.}\ \bibnamefont {Homes}},
  \bibinfo {author} {\bibfnamefont {D.~N.}\ \bibnamefont {Basov}}, \ and\
  \bibinfo {author} {\bibfnamefont {S.~G.}\ \bibnamefont {Sharapov}},\
  }\bibfield  {title} {\enquote {\bibinfo {title} {Three-dimensional dirac
  fermions in quasicrystals as seen via optical conductivity},}\ }\href
  {\doibase 10.1103/PhysRevB.87.235121} {\bibfield  {journal} {\bibinfo
  {journal} {Phys. Rev. B}\ }\textbf {\bibinfo {volume} {87}},\ \bibinfo
  {pages} {235121} (\bibinfo {year} {2013})}\BibitemShut {NoStop}%
\bibitem [{\citenamefont {Ashby}\ and\ \citenamefont
  {Carbotte}(2014)}]{Ashby14}%
  \BibitemOpen
  \bibfield  {author} {\bibinfo {author} {\bibfnamefont {P.~E.~C.}\
  \bibnamefont {Ashby}}\ and\ \bibinfo {author} {\bibfnamefont {J.~P.}\
  \bibnamefont {Carbotte}},\ }\bibfield  {title} {\enquote {\bibinfo {title}
  {Chiral anomaly and optical absorption in weyl semimetals},}\ }\href
  {\doibase 10.1103/PhysRevB.89.245121} {\bibfield  {journal} {\bibinfo
  {journal} {Phys. Rev. B}\ }\textbf {\bibinfo {volume} {89}},\ \bibinfo
  {pages} {245121} (\bibinfo {year} {2014})}\BibitemShut {NoStop}%
\bibitem [{\citenamefont {Xu}\ \emph {et~al.}(2016)\citenamefont {Xu},
  \citenamefont {Dai}, \citenamefont {Zhao}, \citenamefont {Wang},
  \citenamefont {Yang}, \citenamefont {Zhang}, \citenamefont {Liu},
  \citenamefont {Xiao}, \citenamefont {Chen}, \citenamefont {Taylor},
  \citenamefont {Yarotski}, \citenamefont {Prasankumar},\ and\ \citenamefont
  {Qiu}}]{Xu16}%
  \BibitemOpen
  \bibfield  {author} {\bibinfo {author} {\bibfnamefont {B.}~\bibnamefont
  {Xu}}, \bibinfo {author} {\bibfnamefont {Y.~M.}\ \bibnamefont {Dai}},
  \bibinfo {author} {\bibfnamefont {L.~X.}\ \bibnamefont {Zhao}}, \bibinfo
  {author} {\bibfnamefont {K.}~\bibnamefont {Wang}}, \bibinfo {author}
  {\bibfnamefont {R.}~\bibnamefont {Yang}}, \bibinfo {author} {\bibfnamefont
  {W.}~\bibnamefont {Zhang}}, \bibinfo {author} {\bibfnamefont {J.~Y.}\
  \bibnamefont {Liu}}, \bibinfo {author} {\bibfnamefont {H.}~\bibnamefont
  {Xiao}}, \bibinfo {author} {\bibfnamefont {G.~F.}\ \bibnamefont {Chen}},
  \bibinfo {author} {\bibfnamefont {A.~J.}\ \bibnamefont {Taylor}}, \bibinfo
  {author} {\bibfnamefont {D.~A.}\ \bibnamefont {Yarotski}}, \bibinfo {author}
  {\bibfnamefont {R.~P.}\ \bibnamefont {Prasankumar}}, \ and\ \bibinfo {author}
  {\bibfnamefont {X.~G.}\ \bibnamefont {Qiu}},\ }\bibfield  {title} {\enquote
  {\bibinfo {title} {Optical spectroscopy of the weyl semimetal taas},}\ }\href
  {\doibase 10.1103/PhysRevB.93.121110} {\bibfield  {journal} {\bibinfo
  {journal} {Phys. Rev. B}\ }\textbf {\bibinfo {volume} {93}},\ \bibinfo
  {pages} {121110} (\bibinfo {year} {2016})}\BibitemShut {NoStop}%
\bibitem [{\citenamefont {de~Sousa}\ and\ \citenamefont
  {Chen}(2023)}]{deSousa23_graphene_opacity}%
  \BibitemOpen
  \bibfield  {author} {\bibinfo {author} {\bibfnamefont {M.~S.~M.}\
  \bibnamefont {de~Sousa}}\ and\ \bibinfo {author} {\bibfnamefont
  {W.}~\bibnamefont {Chen}},\ }\bibfield  {title} {\enquote {\bibinfo {title}
  {Opacity of graphene independent of light frequency and polarization due to
  the topological charge of the dirac points},}\ }\href {\doibase
  10.1103/PhysRevB.108.165201} {\bibfield  {journal} {\bibinfo  {journal}
  {Phys. Rev. B}\ }\textbf {\bibinfo {volume} {108}},\ \bibinfo {pages}
  {165201} (\bibinfo {year} {2023})}\BibitemShut {NoStop}%
\end{thebibliography}%

\end{document}